\documentclass[aps,prb,preprint,superscriptaddress]{revtex4-1}

\usepackage{graphicx}
\usepackage{dcolumn}
\usepackage{bm}
\usepackage{amssymb}
\usepackage{comment}
\usepackage{tikz}
\usetikzlibrary{scopes}
\usetikzlibrary{calc}
\usepackage{amsthm}
\usepackage{amsopn}
\usepackage{amsmath}
\usepackage[caption=false]{subfig}
\usepackage{braket}

\newcommand{\ve}[1]{\boldsymbol{#1}}

\newcommand{\mro}{\mathrm{odd}}

\begin{document}

\title{A reexamination of the effective fine structure constant of graphene, as measured in graphite}%

\author{Yu Gan}
\affiliation{Department of Physics and Federick Seitz Materials Research Laboratory, University of Illinois, Urbana, IL 61801}
\author{Gilberto de la Pena Munoz}
\affiliation{Department of Physics and Federick Seitz Materials Research Laboratory, University of Illinois, Urbana, IL 61801}
\author{Anshul Kogar}
\affiliation{Department of Physics and Federick Seitz Materials Research Laboratory, University of Illinois, Urbana, IL 61801}
\author{Bruno Uchoa}
\affiliation{University of Oklahoma, Norman, OK 73019}
\author{Diego Casa}
\affiliation{Advanced Photon Source, Argonne National Laboratory, Argonne, IL 60439}
\author{Thomas Gog}
\affiliation{Advanced Photon Source, Argonne National Laboratory, Argonne, IL 60439}
\author{Eduardo Fradkin}
\affiliation{Department of Physics and Federick Seitz Materials Research Laboratory, University of Illinois, Urbana, IL 61801}
\author{Peter Abbamonte}
\affiliation{Department of Physics and Federick Seitz Materials Research Laboratory, University of Illinois, Urbana, IL 61801}
\affiliation{Advanced Photon Source, Argonne National Laboratory, Argonne, IL 60439}

\begin{abstract}
We present a refined and improved study of the influence of screening on the effective fine structure constant of graphene, $\alpha^\ast$, as measured in graphite using inelastic x-ray scattering. This follow-up to our previous study\cite{reed} was carried out with two times better energy resolution, five times better momentum resolution, and improved experimental setup with lower background. We compare our results to RPA calculations and evaluate the relative importance of interlayer hopping, excitonic corrections, and screening from high energy excitations involving the $\sigma$ bands. We find that the static, limiting value of $\alpha^\ast$ falls in the range 0.25 to 0.35, which is higher than our previous result of 0.14\cite{reed}, but still below the value expected from RPA. We show the reduced value is not a consequence of interlayer hopping effects, which were ignored in our previous analysis\cite{reed}, but of a combination of excitonic effects in the $\pi \rightarrow \pi^*$ particle-hole continuum, and background screening from the $\sigma$-bonded electrons. We find that $\sigma$-band screening is extremely strong at distances of less than a few nm, and should be highly effective at screening out short-distance, Hubbard-like interactions in graphene as well as other carbon allotropes. 
\end{abstract}

\maketitle

\section{\label{sec:intro}Introduction}

Since its synthesis a decade ago, graphene has been the subject of intense research\cite{graphenereview,transportrev}.  In the long-wavelength limit, graphene satisfies the massless Dirac equation, where its electrons follow a linear band dispersion given by $E = \pm \hbar v_f q$, where the $\pm$ refers to the conduction/valence bands, $v_f \sim 10^8 \mathrm{cm/s} \approx c/300$ is the Fermi velocity, and $q$ is the electron momentum (in 2D)\cite{wallace,dirac}.  As a result of this dispersion, the bare electron-electron interaction strength -- as defined by the fine-structure constant, $\alpha = U/K \approx 2.2/\epsilon$, where $\epsilon$ is the dielectric constant of the substrate on which the graphene is mounted ($\alpha \approx 2.2$ for suspended graphene) -- is of order unity, in contrast to the interaction strength in QED, where $\alpha = 1/137$; thus, many-body effects should, in principle, play a significant role in graphene\cite{kotov}.  In particular, the screened value of $\alpha$, which one could call a renormalized or effective fine structure constant, $\alpha^\ast$, is a subject of great theoretical interest\cite{ulybyshev,fogler,kotov,non-fermi_renorm,mishchenko}, and is relevant to all Dirac systems including topological insulator surface states\cite{kane,qi}, some classes of transition metal dichalcogenides\cite{rossnagel}, three-dimensional Dirac and Weyl semimetals\cite{ashvin,young}, etc. 

Experiment and theory, however, give conflicting views on the value of $\alpha^\ast$.  Using inelastic X-ray scattering (IXS) experiments performed on graphite\cite{reed}, we previously found that, for freestanding graphene, $\alpha^\ast \approx 0.15$ as $q \rightarrow 0$ and $\omega \rightarrow 0$, and cited excitonic shifts in the $\pi \rightarrow \pi^*$, particle-hole continuum as the origin of the reduced value of $\alpha^\ast$. Interpretation of this experiment was based on the assumption that the interaction between graphene sheets in graphite is primarily Coulombic, with interlayer hopping playing a secondary role\cite{reed}. The experiment, however, had an energy resolution of only 0.3 eV, when the features of primary interest range from 1 - 15 eV.  Additionally, the lowest momenta measured were at 0.238 \AA$^{-1}$ and 0.476 \AA$^{-1}$ so the extrapolation of $q \rightarrow 0$ relied on only a few points.  Subsequent calculations in the random-phase-approximation\cite{katsnelson} (RPA) using the full $\pi$-bands and including the interlayer hopping parameters $\gamma_3$ and $\gamma_1$ in graphite (see Fig. \ref{fig:structure}) suggested that this discrepancy was due to graphitic effects, i.e., the presence of interlayer hopping effects ignored in our analysis. 


In this paper, we present a refined measurement of $\alpha^\ast$ in graphene as measured in graphite, using an improved version of the previous method\cite{reed}, with higher energy and momentum resolution, as well as reduced elastic background.  We evaluate the importance of interlayer hopping, excitonic corrections, and the higher energy $\sigma$-bands to the effective fine-structure constant.  We compare to theoretical results obtained within the RPA, and show that for $q \sim 0.212$ \AA$^{-1}$ (length scale $\sim$ 30 \AA), $\alpha^\ast$ falls in the range 0.25 to 0.35, which is well below the RPA value for graphene. We show the origin of this anomalously low value is not graphitic (i.e., interlayer hopping) effects, but a combination of screening from the $\sigma$-bonded electrons, which contribute significantly at finite momentum, and excitonic effects in the $\pi$ bands, which comprise a beyond-PRA correction to the screening. While the effect of $\sigma$ band screening decreases with decreasing $q$, we argue such effects remain relevant to length scales up to $\sim 30 \AA$. 

\section{\label{sec:experiment}Experiment}

IXS measurements were carried out in Sector 9 at the Advanced Photon Source. Energy analysis was done with a diced, Si-444, spherical backscattering analyzer operating at 7.81 keV, which imaged scattered photons onto a MYTHEN microstrip detector. The total energy resolution of the instrument was 175 meV, which is a factor of two better than our previous study\cite{reed}. 

For these measurements, we constructed a new sample chamber equipped with a moving, in-vacuum beam stop, providing access to scattering angles as low as $1^{\circ}$. The momentum resolution was set to 0.15-0.3 \AA$^{-1}$, depending upon the scan region. The chamber was designed for windowless operation with the synchrotron beam pipes, reducing scattering from upstream vacuum windows. The final setup exhibited greatly reduced background in the 1-5 eV range, particularly from the quasielastic line, a key source of error as described in Appendix \ref{app:sumfit}. 

Experiments were done on both ZYA-grade, highly-ordered pyrolitic graphite (HOPG) and high quality, commercially obtained graphite single-crystals. The spectra of the two were found to be indistinguishable in the momentum range in this study. The experimental data were placed in absolute units of $(eV \AA^3)^{-1}$ by normalizing using the $F$ sum rule, as described in Ref. \cite{reed} and Appendix \ref{app:sumfit}. 

\section{\label{sec:conversion}Conversion of Experimental Data}

In order to obtain information about $\alpha^\ast$, we need to determine the two-dimensional, density-density response function of graphene, $\chi_{2D}(\ve{q},\omega)$, which is the most general description of the collective charge dynamics of the system\cite{reed}. IXS measurements cannot be carried out on single-layer graphene, however, so a means of determining properties of graphene from measurements on graphite is required. 


The measured intensity in an IXS experiment at a given momentum and energy is proportional to the dynamic structure factor, $S(\ve{q},\omega)$, 
which is related to the density-density response function of three-dimensional graphite, $\chi_{3D}(q,\omega)$, by the quantum mechanical version of the fluctuation-dissipation theorem,

\begin{equation}
\label{eq:f-d}
S(\ve{q},\omega)=-\frac{1}{\pi}\frac{1}{1-e^{-\beta\hbar\omega}}\operatorname{Im}\chi_{3D}(\ve{q},\omega).
\end{equation}
We note that $\chi_{3D}$ is causal and satisfies the Kramers-Kronig relation
\begin{equation}
	\operatorname{Re}\chi_{3D}(\ve{q},\omega)=
		\frac{2}{\pi}\mathcal{P}\int_0^{\infty}{d\omega' 
		\frac{\omega'\operatorname{Im}\chi_{3D}(\ve{q},\omega')}{(\omega')^2-\omega^2}}.
	\label{eq:kramers}
\end{equation}

\begin{figure*}
	\subfloat[][]{
		\includegraphics[width=0.45\textwidth]{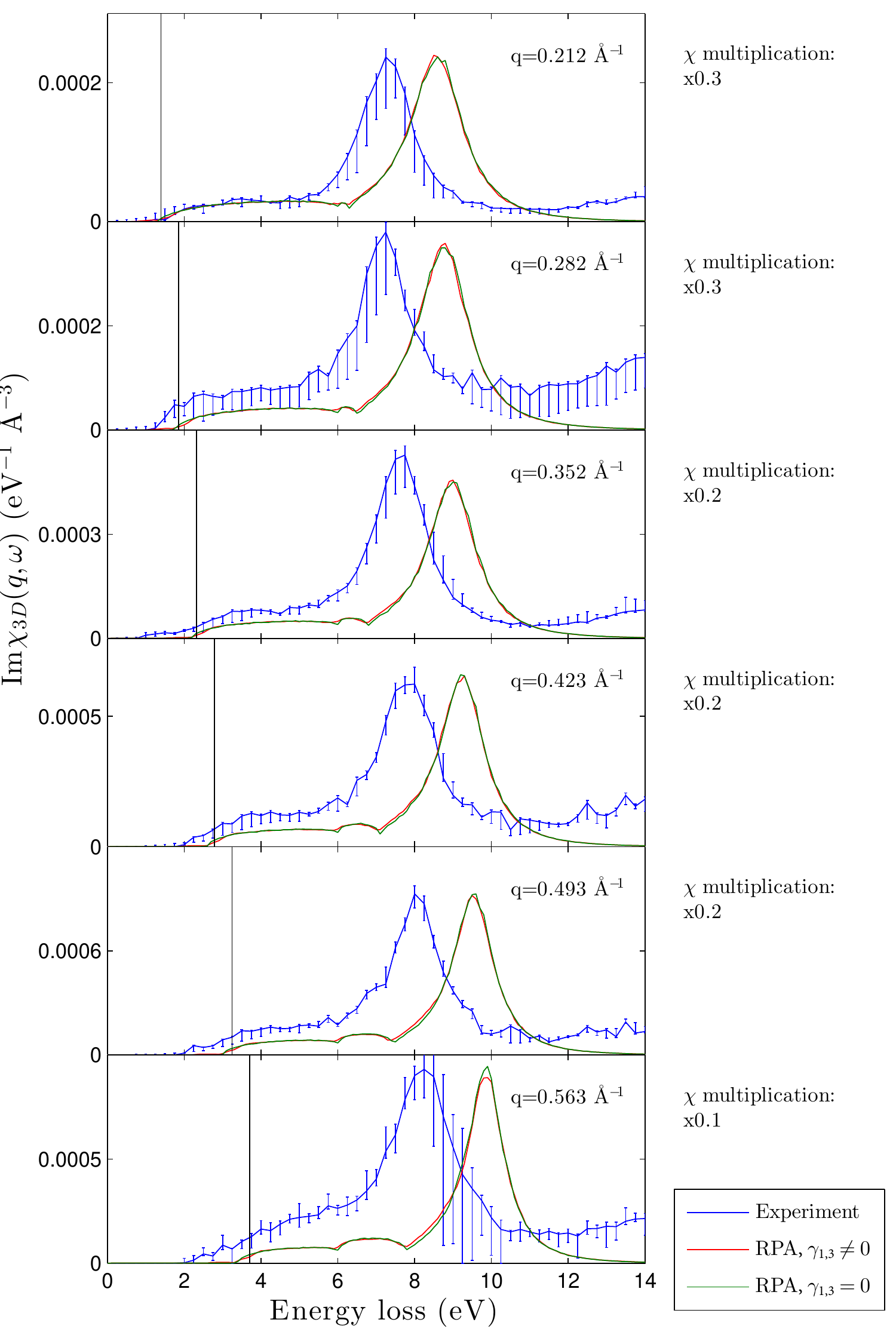}
		\label{fig:imchi3d}
	}
	\subfloat[][]{
		\includegraphics[width=0.45\textwidth]{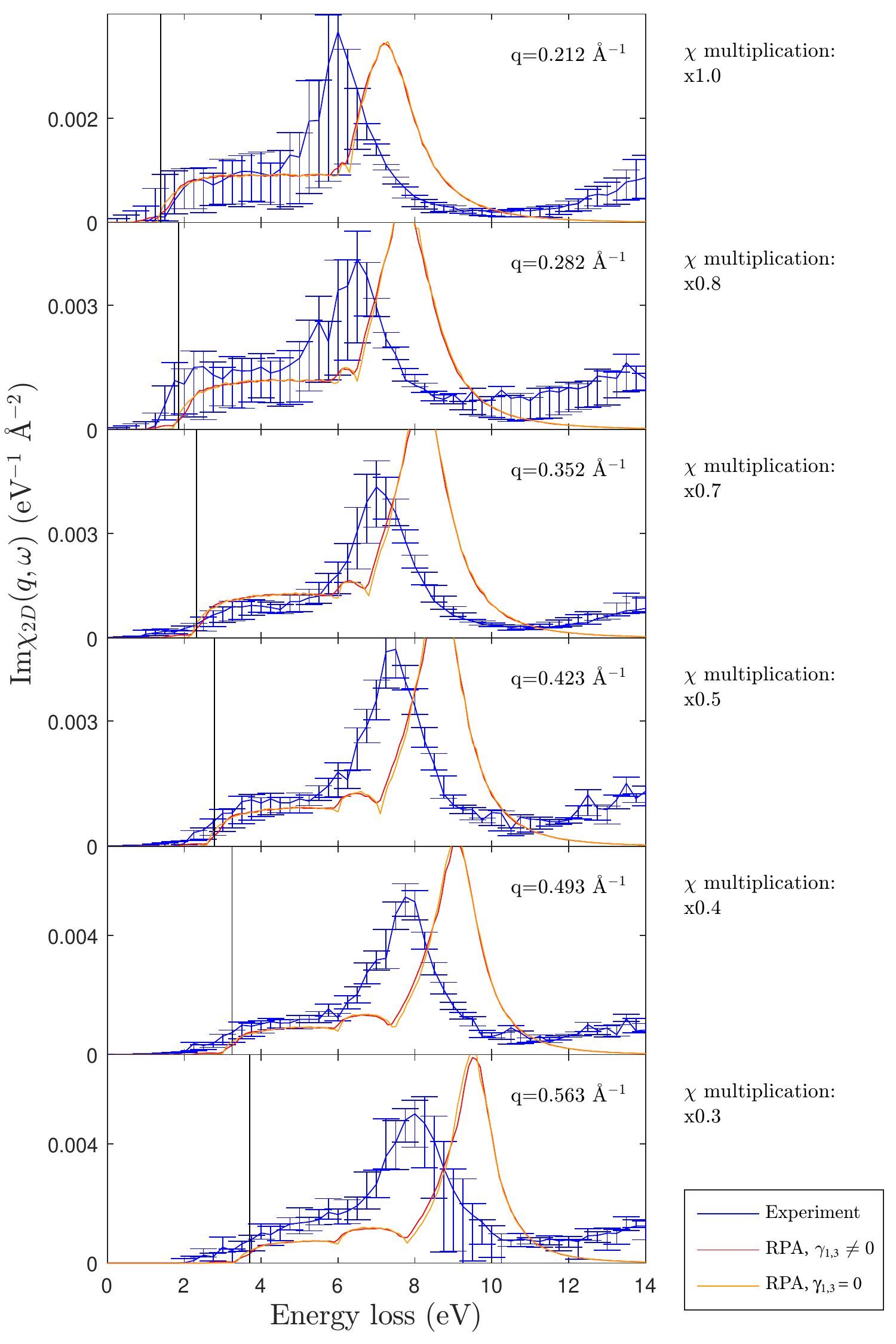}
		\label{fig:imchi2d}
	}
	\caption[]{The imaginary part of the experimental density response of graphite and graphene compared to results from RPA calculations. (a) $-\operatorname{Im}\chi_{3D}(\ve{q},\omega)$ from IXS measurements (blue points), RPA calculations including interlayer hopping (green line) and omitting interlayer hopping (red line), plotted against energy for selected in-plane momenta, $q$. The RPA curves had to be multiplied by an arbitrary scale factor to allow visual comparison to the experiment. The black, vertical lines indicate the energy, $\hbar v_F q$, at which the onset of the particle-hole continuum is expected in RPA. (b) $-\operatorname{Im}\chi_{2D}(\ve{q},\omega)$ determined from IXS measurements from Eq. 3 (blue points), RPA calculations including interlayer hopping (red line) and omitting interlayer hopping (orange line), plotted against energy for selected in-plane momenta, $q$. The black line, again, indicates the onset energy of the continuum expected from RPA. Experimental error bars are derived from both Poisson statistics and our fitting parameters (see Appendix \ref{app:sumfit}).}
	\label{fig:chiplots}
\end{figure*}

\noindent
So, by extrapolating to $\omega \rightarrow \infty$, interpolating the discrete points in $\omega$, and integrating, it is possible to determine the full, complex $\chi_{3D}(\ve{q},\omega)$ for graphite in absolute units of $(eV \AA^3)^{-1}$, as shown in Fig. \ref{fig:imchi3d}.  Note that the onset of $\pi \rightarrow \pi^*$ particle-hole excitations is visible in the spectra near $\omega= v_f q$, indicating that we are probing the relevant, low-energy valence excitations. The peak at $\omega \approx 8$ eV is the well-known $\pi$-plasmon, which is not a free carrier plasmon but an effect of the Van Hove singularity at the top of the $\pi$ band\cite{reed}.

Learning about graphene requires finding a relationship between its density response, $\chi_{2D}(\ve{q},\omega)$, and the measured, three-dimensional response of graphite, $\chi_{3D}(\ve{q},\omega)$. This can be accomplished by making three assumptions: (1) that the primary interaction between layers in graphite is electrostatic, with interlayer hopping playing a negligible role, (2) that the layers are arbitrarily thin, and (3) that the electrons are distributed homogeneously within each layer. In this case, the two response functions are related by \cite{reed}

\begin{equation}
	\chi_{2D}(\ve{q},\omega) = \frac{\chi_{3D}(\ve{q},\omega) d}{1 - V_{2D}(\ve{q})[1-F(\ve{q})]\chi_{3D}(\ve{q},\omega) d}.
	\label{eq:2d3d}
\end{equation}

\noindent
Here, $V_{2D}(\ve{q}) = 2\pi e^2 /q$ is the two-dimensional Coulomb interaction, and $F(\ve{q}) = \sinh(q d)/(\cosh(q d)-\cos(q_z d))$ is a structure factor that describes the Coulomb interaction in a layered, three-dimensional system, i.e., $V_{3D}(\ve{q}) = V_{2D}(\ve{q}) F(\ve{q}) d$, where $d=3.35\AA$ is the interlayer spacing\cite{shung,reed,kotov}.

Eq. \ref{eq:2d3d} makes use of the fact that, if the above three assumptions are valid, the polarization function of graphene, $\Pi_{2D}(\ve{q},\omega)$, is (apart from units) identical to that for graphite,\cite{reed} i.e., 

\begin{equation}
	\Pi_{2D}(\ve{q},\omega) = \Pi_{3D}(\ve{q},\omega) d.
	\label{eq:pi2d3d}
\end{equation}

\noindent
These quantities are related to the respective susceptibilities by $\chi_{3D}=\Pi_{3D}/\epsilon_{3D}$ and $\chi_{2D}=\Pi_{2D}/\epsilon_{2D}$, where $\epsilon_{3D}=1-V_{3D}\Pi_{3D}$ and $\epsilon_{2D}=1-V_{2D}\Pi_{2D}$ are the dielectric functions.  The spectra for $\operatorname{Im}\chi_{2D}$ obtained in this manner are shown in Fig. \ref{fig:imchi2d}, and the associated polarization functions in Fig. \ref{fig:piplots}. Crucially, the $\chi_{2D}(\ve{q},\omega)$ obtained from Eq. \ref{eq:2d3d} exhibits the correct asymptotic properties expected for two-dimensional graphene (see Fig. \ref{fig:chi0} and Section VI). 


Written in terms of the other quantities, the screened, effective fine structure constant of graphene $\alpha^\ast$ is

\begin{equation}
	\alpha^\ast(\ve{q},\omega) = \frac{\alpha}{\epsilon_{2D}} = \alpha \cdot \left[1 + V_{2D}(\ve{q})\chi_{2D}(\ve{q},\omega)\right].
	\label{eq:alphastar}
\end{equation}

\noindent
This quantity can be thought of as the value of the fine structure constant incorporating screening corrections to all orders in perturbation theory. Note that, although Eq. \ref{eq:alphastar} appears similar to an RPA expression, we have at no point assumed that only RPA polarization bubbles contribute to the susceptibility; indeed, the $\chi$ we recover from experiment includes all screening processes including excitonic effects and other corrections beyond the RPA.  


\section{\label{sec:rpasect}RPA Calculations}

\begin{figure}
	\subfloat[][]{
		\includegraphics[width=0.35\columnwidth]{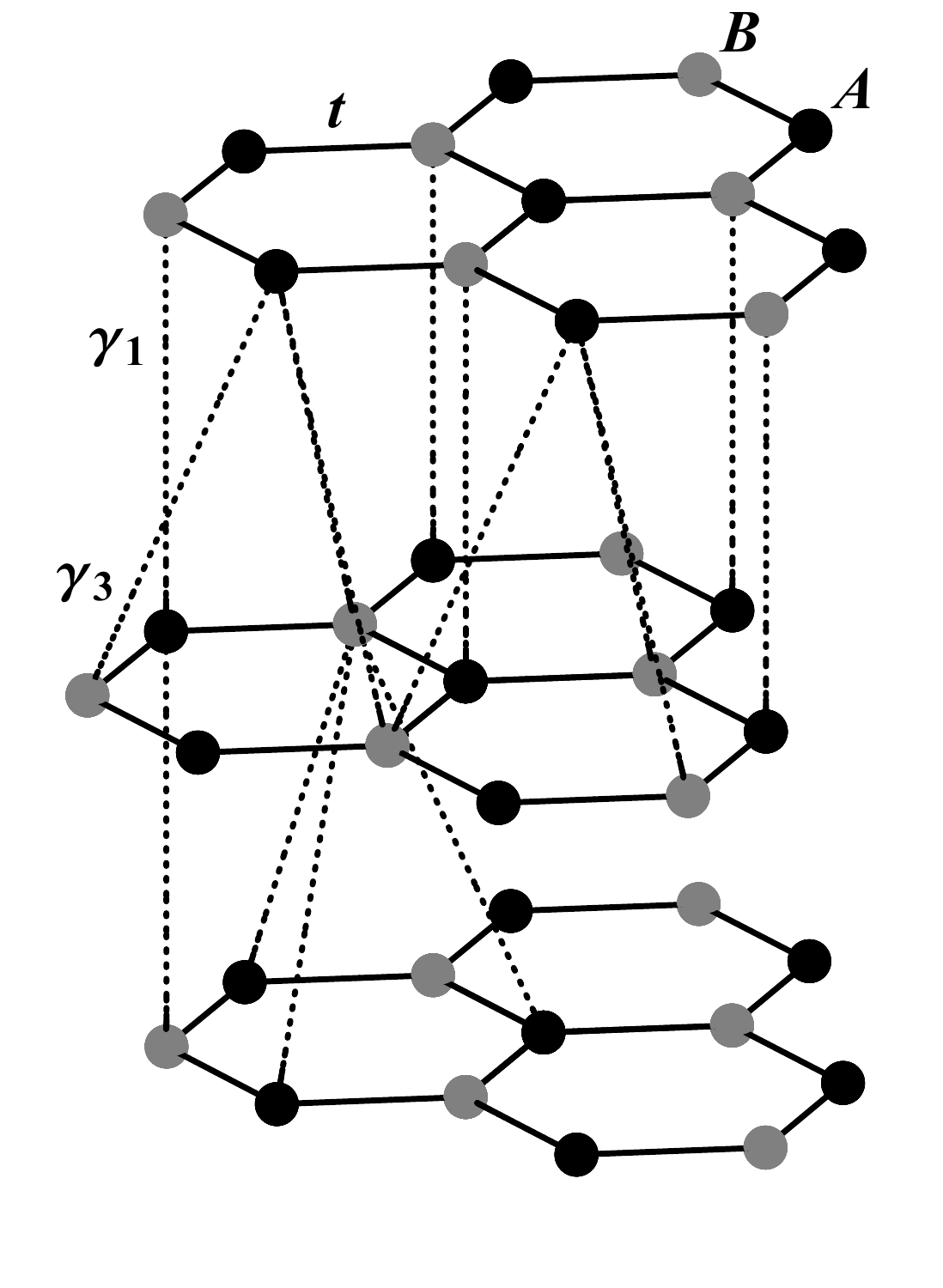}
		\label{fig:structure}
	}
	\subfloat[][]{
		\includegraphics[width=0.55\columnwidth]{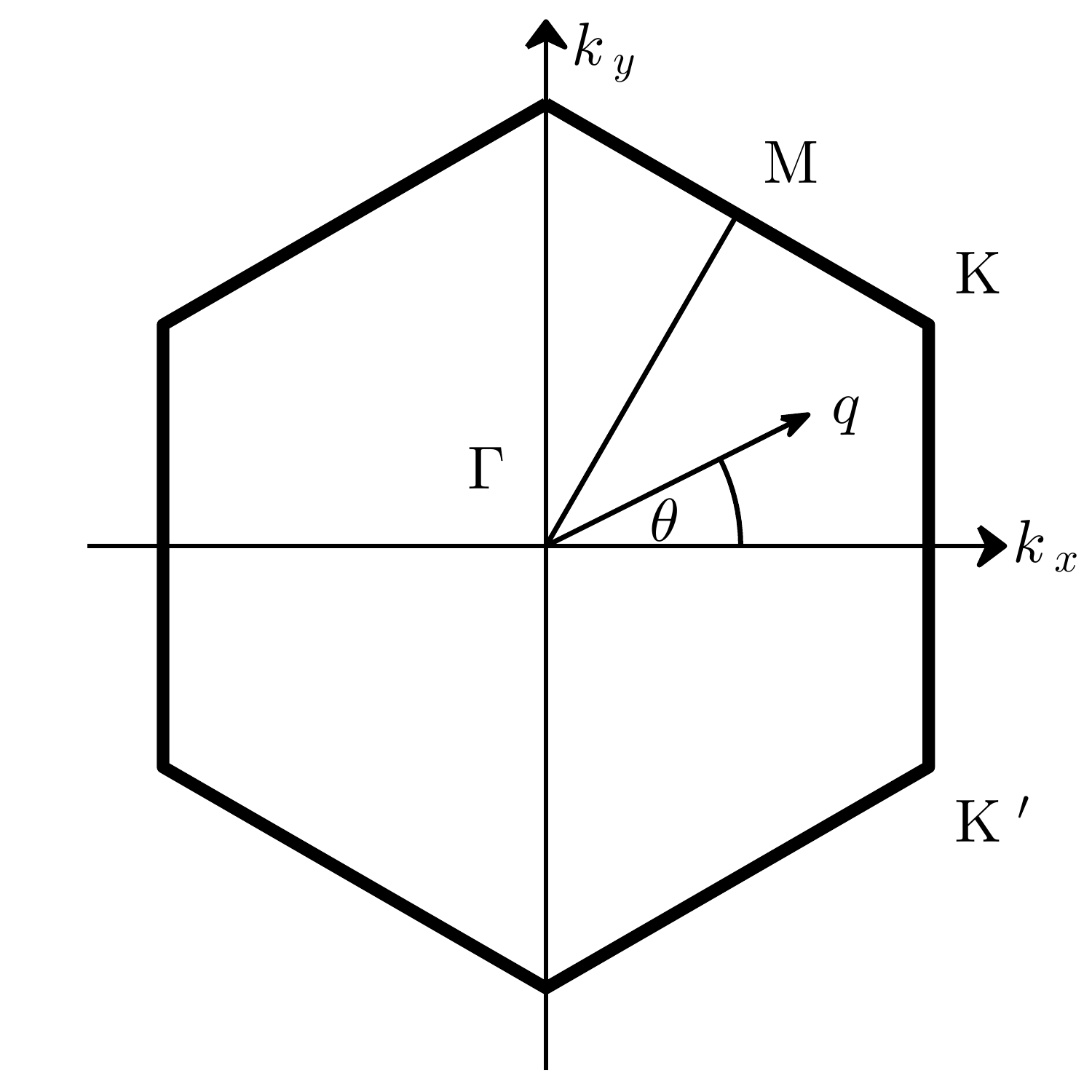}
		\label{fig:brillouin}
	}
	\caption[]{\subref{fig:structure} Crystal structure of ABA-stacked graphite defining the hopping parameters $t$, $\gamma_1$, and
			$\gamma_3$. \subref{fig:brillouin} Brillouin zone of graphene defining the in-plane azimuthal angle of the basal plane. The spectra were found to be independent of this angle in the range of momenta measured in this study.}
\end{figure}

To aid interpretation of the experimental data, we performed a $\pi$-electron tight binding calculation of the susceptibilities of graphite and graphene in the random phase approximation (RPA), following the approach of Yuan et al. \cite{katsnelson}. In this approach, one calculates the polarization functions for single-layer graphene (SLG), $\Pi_{2D}$, and multilayer graphite (MLG), $\Pi_{3D}$, using the Lindhard formula, as described in Appendix B. The polarization functions are then used to determine the dielectric functions and susceptibilities via the relationships discussed in Section III. 

The RPA approach neglects excitonic effects, which have been argued to be crucial for interpreting the particle-hole excitation spectrum of graphene\cite{reed,makHeinzExc}. However, RPA has the advantage of allowing one to switch on and off the interlayer hopping parameters, $\gamma_1$ and $\gamma_3$ defined in Fig. \ref{fig:structure}, allowing one to evaluate the consequences of both interlayer hopping and excitonic effects by comparing to the experimental IXS data. 

Like Yuan \cite{katsnelson}, our calculations are based on a tight binding theory that includes only the electrons in the $\pi$ bands (see Appendix B). The effects of the deeper $\sigma$ bands are incorporated phenomenologically as a frequency-independent, background dielectric constant, \citep{kappakats,kappaorig}

\begin{equation}
	\kappa_N(\ve{q}) = \frac{\kappa_0 + 1 - (\kappa_0 -1)e^{-q L_N}}{\kappa_0 + 1 + (\kappa_0-1)e^{-q L_N}}\kappa_0
	\label{eq:kappa}
\end{equation}

\noindent
where $\kappa_0 = 2.4$ is the background dielectric constant of graphite\cite{dresselhaus}, $L_N=d_m + (N - 1) d$ is the height of a multilayer graphite system with $N$ layers, and $d_m = 2.8 \AA$ is the thickness of a single graphene sheet \cite{katsnelson}. Eq. \ref{eq:kappa} exhibits the proper scaling with the number of layers at low momenta, and was argued to correctly describe screening by excitations involving the sigma bands\cite{katsnelson}. Note that for infinite, MLG, this expression reduces to $\kappa_\infty(\ve{q})=\kappa_0$ for any nonzero momentum. 

We differ from Yuan\cite{katsnelson}, however, in our use of $\kappa_N(\ve{q})$. In Ref. \cite{katsnelson}, $\kappa_N(\ve{q})$ was used only to reduce the strength of the Coulomb interaction, by using a screened $W(\ve{q})=V(\ve{q})/\kappa(\ve{q})$. As shown in Appendix C, this usage neglects interference between screening by $\pi$ and $\sigma$ electrons. The correct use of $\kappa_N(\ve{q})$ is in the relationship between the polarization and dielectric functions,

\begin{equation}
	\epsilon(\ve{q},\omega) = \kappa_N(\ve{q}) - V(\ve{q})\Pi(\ve{q},\omega),
	\label{eq:epsilon}
\end{equation}

\noindent
where $N$ is chosen to match the dimensionality of the system. 

Having determined the polarization functions for MLG, $\Pi_{3D}(\ve{q},\omega)$, and SLG, $\Pi_{2D}(\ve{q},\omega)$, we calculated four different susceptibilities for comparison to experiment. The first,


\begin{equation}
	\chi_{3D}^{RPA}(\ve{q},\omega) = \frac{\Pi_{3D}(\ve{q},\omega)}{\kappa_\infty(\ve{q}) - V_{2D}(\ve{q})F(\ve{q}) d \, \Pi_{3D}(\ve{q},\omega)},
\end{equation}

\noindent 
is the three-dimensional susceptibility for MLG in the RPA, including the effects of both interlayer hopping and electrostatic coupling between the layers. The imaginary part of this quantity could be compared directly to the experimental data to assess the importance of excitonic effects. The second,

\begin{equation}
	\chi_{2D}^{RPA}(\ve{q},\omega) = \frac{\Pi_{2D}(\ve{q},\omega)}{\kappa_1(\ve{q}) - V_{2D}(\ve{q})\Pi_{2D}(\ve{q},\omega)}
\end{equation}

\noindent
is the two-dimensional susceptibility for SLG. This quantity could be compared to the converted experimental data resulting from the application of Eq. 3 to assess the combined importance of excitonic and interlayer hopping effects. The third quantity 

\begin{equation}
	\tilde{\chi}_{2D}^{RPA}(\ve{q},\omega) = \frac{\Pi_{3D}(\ve{q},\omega) d}{\kappa_1(\ve{q}) - V_{2D}(\ve{q})\Pi_{3D}(\ve{q},\omega) d}
\end{equation}

\noindent
is a two-dimensional propagator describing, in RPA, how the physics of graphene would be modified if $\gamma_{1,3} \neq 0$, i.e., if it were subjected to the same interlayer hopping effects as graphite. Finally, 

\begin{equation}
	\tilde{\chi}_{3D}^{RPA}(\ve{q},\omega) = \frac{\Pi_{2D}(\ve{q},\omega)/d}{\kappa_\infty(\ve{q}) - V_{2D}(\ve{q})F(\ve{q})\Pi_{2D}(\ve{q},\omega)},
\end{equation}

\noindent
is a three-dimensional propagator for graphite in which the interlayer hopping is switched off, i.e., $\gamma_{1,3} = 0$, but the layers are still coupled by electrostatic effects. By comparing these quantites to one another, to the experimental data, and to the converted data determined from Eq. 3, it should be possible to disentangle the contribution of excitonic effects, interlayer hopping, and $\sigma$-band screening on the effective fine structure constant of graphene.

\section{Results}
\label{sec:results}

We are now in a position to evaluate the source of the anomalous screening observed in our original study\cite{reed}. In particular, we wish to know whether this screening is due to the graphitic effects discussed recently\cite{katsnelson}, i.e. the interlayer hopping parameters $\gamma_1$ and $\gamma_3$, background screening from the $\sigma$-bands, or excitonic effects as we claimed previously\cite{reed}.  

In Fig. \ref{fig:chiplots}a we compare the experimental data to the imaginary part of $\chi_{3D}$ computed from RPA both with and without interlayer hopping (Eqs. 8 and 11, respectively). The first thing to note is that RPA does a poor job at reproducing the energy of the so-called ``$\pi$-plasmon" at $\sim 8$ eV. Note that this excitation is not a free carrier plasmon, but a result of the Van Hove singularity at the top of the $\pi$ band\cite{reed}. We attempted to improve the agreement by adjusting the parameters $t$, $\gamma_{1,3}$, and $\kappa_0$, but any parameters that produced the correct plasmon dispersion were dramatically different from the commonly accepted values\cite{kotov,taft} of $t=3$ eV, $\gamma_1 = 0.4$ eV, $\gamma_3 =0.3$ eV, and $\kappa_0 = 2.4$.  Changes to these parameters also drastically distorted the qualitative shape of the $\pi \rightarrow \pi^\ast$ continuum. The discrepancy is most likely due to excitonic effects not captured by RPA\cite{reed,MakHeinzExc}. We therefore show only the spectra calculated using the commonly accepted values for the hopping parameters and $\kappa_0$. Note that, in agreement with Ref. \cite{katsnelson}, the effect of a nonzero $\gamma_{1,3}$ is not to qualitatively change the spectra, but to smear some of the features near threshold. Such effects are subtle and not easily distinguished from broadening due to experimental resolution. 

In addition, a significant discrepancy is observed between the {\it size} of the experimental response and that computed with RPA (Fig. 1a). The latter is significantly larger in magnitude and had to be multiplied by an arbitrary scale factor, ranging from 0.15 to unity, to allow visual comparison with the experiment. This magnitude discrepancy is fundamental, and arises from the significant spectral weight in the $\sigma$-bands in the experimental data that is absent from the RPA result. Despite residing at $\sim$ 25 eV, the $\sigma$-bands are quite polarizable and have significant influence on the value of $\chi$ at low energy, because $\chi$ is a nonlinear function of $\Pi$\cite{taft,mari}. In other words, our RPA approach treats the system as if all the screening is accomplished by the $\pi$ electrons, while in reality some of the work is done instead by the $\sigma$-bonded electrons. RPA therefore predicts that the response from the $\pi$ electrons is anomalously large, since the added $\sigma$ screening channel is absent. The influence of the $\sigma$-bands increases with increasing $\ve{q}$, reaching a maximum at around 2 $\AA^{-1}$. Note that this problem is not corrected by including a background $\kappa_N(q)$, which does not replace spectral weight that is missing from the $F$ sum rule (note that using $\kappa_N(q)$ inside a screened $W(q)$, as was done in Ref. \cite{katsnelson}, does not improve the situation). 

\begin{figure}
	\includegraphics[width=0.7\columnwidth]{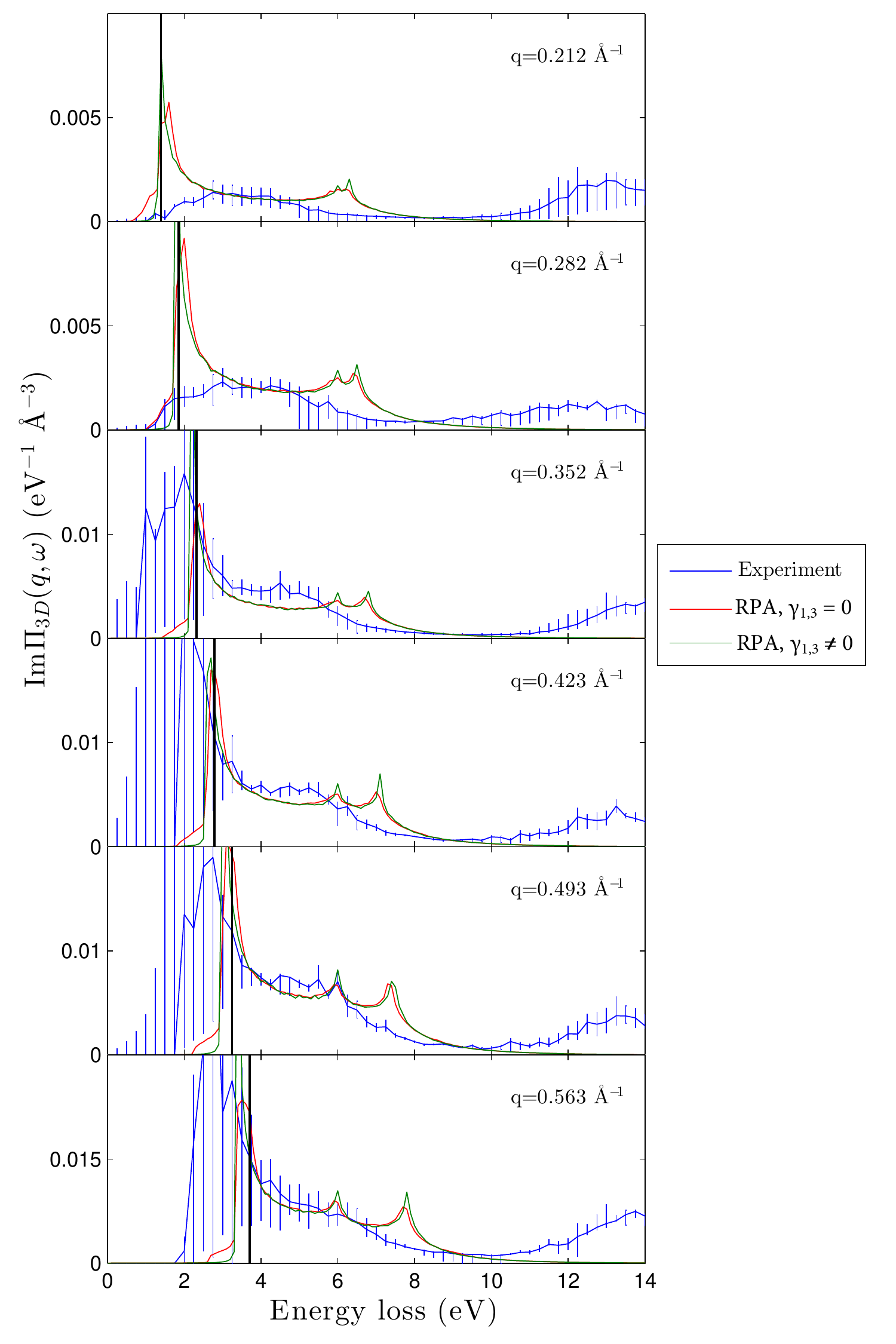}
	\caption[]{Imaginary part of the polarization function of graphite, $-\operatorname{Im}\Pi_{3D}(\ve{q},\omega)$, from experiment (blue points), compared to the results of RPA calculations both with (green line) and without (red line) interlayer hopping, $\gamma_{1,3}$. Within the assumptions described in Section III, $\operatorname{Im}\Pi_{3D}(\ve{q},\omega)$ should be the same as $\operatorname{Im}\Pi_{2D}(\ve{q},\omega)$ apart from an overall factor of the interlayer spacing, $d$. Note that the magnitude discrepancy observed in Fig. 1 is not observed here, because polarization functions from different excitations combine additively (Eq. 7)}
	\label{fig:piplots}
\end{figure}

Still looking at Fig. 1a, another key observation is that the onset of the $\pi \rightarrow \pi^*$ continuum in the experiment is consistently lower in energy than in the RPA calculation. These shifts are even better visible in the imaginary part of the polarization function (Fig. 3), which is better reflective of single particle excitations than the susceptibility, which emphasizes the collective modes\cite{caigraphite}. For momenta $q > 0.35 \AA^{-1}$, these shifts are as much as 1 eV compared to RPA, which is too large to be explained by graphitic effects, and must be attributed to excitonic interaction between the valence hole and photoexcited electron, as discussed previously\cite{reed,MakHeinzExc}. At lower momenta, the shifts are smaller and similar in magnituide to what is expected from interlayer hopping effects, though excitonic effects surely still play some role. 

We turn now to the two-dimensional response, $\operatorname{Im}\chi_{2D}(\ve{q},\omega)$, shown in Fig. 1b, which compares the experimental data converted using Eq. 3 to the two-dimensional RPA response calculated with and without interlayer hopping (Eqs. 9 and 10). The discrepancy in the $\pi$ plasmon energy observed in Fig. 1a, as well as the excitonic shifts in the $\pi \rightarrow \pi^*$ continuum, are visible also in the two-dimensional response. However, the magnitude discrepancy that is so pronounced in three dimensions is much less significant in two dimensions and fades as $q$ decreases. At the lowest momentum point, $q=0.212 \AA^{-1}$, the magnitude discrepancy is absent entirely. This means that in two dimensions, in contrast to the three dimensional case, the polarizability of the high energy $\sigma$ bands ceases to be important in the limit of low momentum. This confirms one's intuition that screening from high energy excitations should be unimportant in two dimensions as $q \rightarrow 0$. 

\section{discussion}

We now turn to the fundamental question of how the effects of interlayer hopping, excitonic shifts in the spectra, and the polarizability of the $\sigma$ bands influence the value of the screened, effective fine structure constant, $\alpha^*(\ve{q},\omega) = \alpha_0 [1 + V_{2D}(\ve{q}) \chi_{2D}(\ve{q},\omega)]$. The quantity $\alpha^*(\ve{q},\omega)$ is a strong function of both energy and momentum, and its value differs greatly depending upon the relative size of the energy, $\omega$, and the momentum, $q v_F$. For transport experiments in a micron-sized device at low temperature, $\omega/v_F q \sim k_B T L / 2 \pi \hbar v_F \sim 10^{-3}$. So the quantity of primary interest is the zero-frequency quantity $\alpha^*(\ve{q},0)$ in the limit of small momentum. We note that this quantity is purely real, and could be either larger or smaller than the bare value, $\alpha_0 = e^2/\epsilon_0 \hbar v_F$. 

\begin{figure*}
	\subfloat[][]{
		\includegraphics[width=0.53\textwidth]{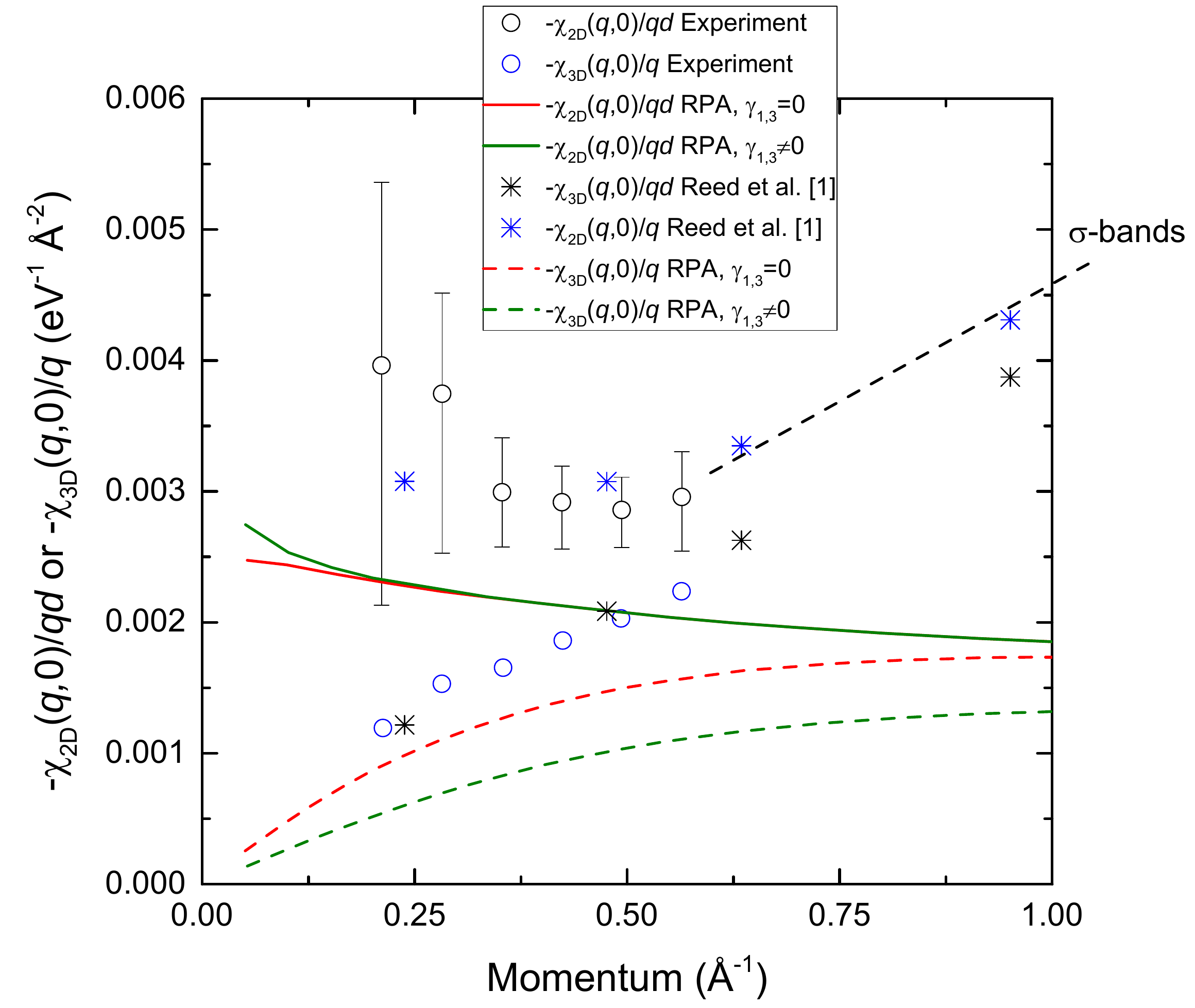}
		\label{fig:chi0}
	}
	\subfloat[][]{
		\includegraphics[width=0.47\textwidth]{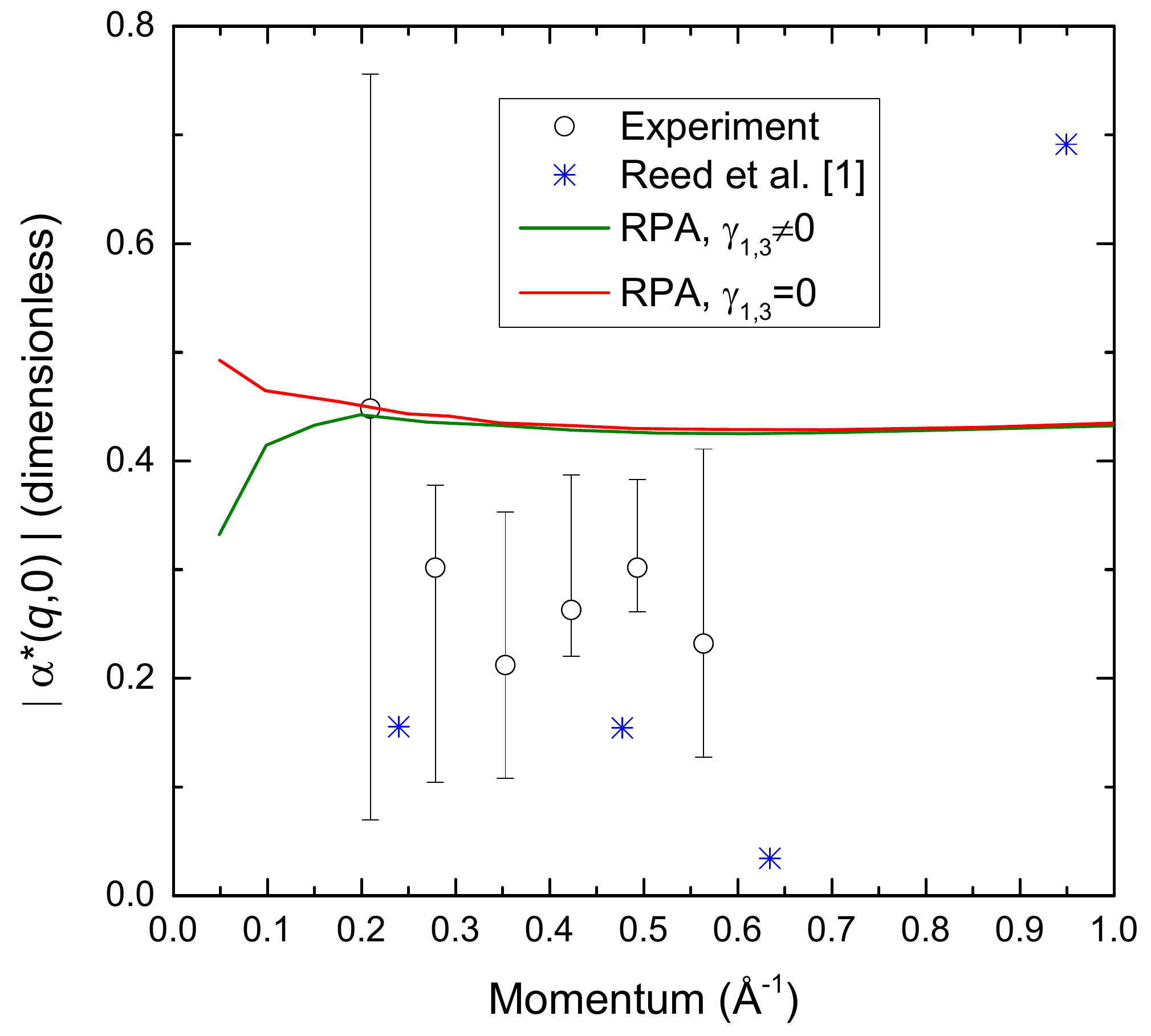}
		\label{fig:alpha0}
	}
	\caption[]{\subref{fig:chi0} Asymptotic screening and the effective fine structure constant of graphene. (a) Combined plot of the static ratios $-\operatorname{Im}\chi_{3D}(q,0)/q$ and $-\operatorname{Im}\chi_{2D}(q,0)/q d$ from experiment and from RPA calculations both with ($\gamma_{1,3} \neq 0$) and without ($\gamma_{1,3} = 0$) interlayer hopping. The results from our previous study\cite{reed} are also plotted, for comparison. This plot shows that all quantities exhibit the proper asymptotic properties in the regime of small $q$. (b) Momentum dependence of the magnitude of the static value of the fine structure constant of graphene. This plot suggests a limiting value in the range 0.25-0.35. 	
	}
	\label{fig:statics}
\end{figure*}

In Fig. 4a, we plot the momentum dependence of the ratios $\chi_{2D}(\ve{q},0)/q d$ and $\chi_{3D}(\ve{q},0)/q$ for both the experimental data as well as the RPA calculations both with ($\gamma_{1,3} \neq 0$) and without ($\gamma_{1,3} = 0$) interlayer hopping. Note that both of these quantities are real and have units $(eV \AA^2)^{-1}$, allowing them to be compared on a single scatter plot. As expected, the quantity $\chi_{3D}(\ve{q},0)/q$ tends toward 0 as $q \rightarrow 0$, since the susceptibility in three dimensions is expected to scale like $q^2$ at small $q$. All three $\chi_{2D}(\ve{q},0)/q d$ curves, on the other hand, converge to a constant as $q \rightarrow 0$, indicating that the quantity $\chi_{2D}(\ve{q},0)$ scales linearly in $q$ at low momenta, which is expected in two dimensions. This scaling is crucially important for two reasons. First, it demonstrates that our conversion expression, Eq. 3, gives the proper asymptotic behavior of the susceptibility in two dimensions, which is a powerful validation of our method. The second is that the product $V_{2D}(\ve{q}) \cdot \chi_{2D}(\ve{q},0)$ converges to a constant at low $q$ for all momenta, indicating that graphene exhibits a finite screening strength, $\epsilon(\ve{q},0) = [1+V_{2D}(\ve{q})\chi_{2D}(\ve{q},0)]^{-1}$, in the long-wavelength limit. 

An important conclusion about interlayer hopping can immediately be reached from Fig. 4a by comparing the RPA calculations of $\chi_{2D}(\ve{q},0)/q d$ with and without interlayer hopping. Notice that the curves are only distinguishable for momenta less than about 0.2 $\AA^{-1}$, which is below the lowest momentum measured in our experiment. What this implies is that, while graphitic effects can in principle influence the value of $\alpha^*$ at low momenta, such effects are likely irrelevant in the regime of our experiment, and that excitonic corrections are in part responsible for deviations compared to RPA visible in Figs. 1a,b at momenta $q < 0.35 \AA^{-1}$. 

Fig. 4a also reveals a major discrepancy between the experiment and theoretical curves at large momenta. For $q > 0.6 \AA^{-1}$, the RPA curves drop off gradually, while the experimental points increase dramatically. This increase is a consequence of screening from the $\sigma$ bands, which exhibits a broad maximum at $q \sim 2 \AA^{-1}$, which is of the order the carbon-carbon bond length\cite{reed}. This discrepancy is, again, a consequence of the absence of $\sigma$ band electrons in the RPA calculation. Hence, the quantity $\kappa(\ve{q})$ taken from Ref. \cite{katsnelson}, while argued to be a good description at small $q$, fails to describe any of the major features of $\sigma$ band screening at larger momenta. Fortunately, the influence of these high energy excitations on $\chi_{2D}$ fades at small momentum, suggesting that we should still get useful information about the low-$q$ value of $\alpha^*$ from our study. 

Finally, in Fig. 4b we plot the magnitude of the static, effective fine structure constant, $\alpha^*(\ve{q},0)$, as a function of momentum for the experiment (i.e., deduced from Eq. 3), and for RPA both with and without interlayer hopping. For reference, the points from our previous study \cite{reed} are also shown. The difference between the two experiments can be attributed to better resolution, better statistics, and lower experimental background in the current study. For momenta $q > 0.5 \AA^{-1}$ the value of $\alpha^*$ is dominated by screening from the $\sigma$ bands. Unfortunately, the error bars on the lowest momentum point, at which $\sigma$-band effects are negligible, are extremely large due to extrapolation of the $F$ sum rule (see Appendix A). The remaining points are converging toward a value in the range of 0.25 to 0.35. This quantity is larger than given in our previous study, but still smaller than the value expected from RPA, and implies an asymptotic dielectric constant in the range 6.2 to 8.8. Because interlayer hopping effects are irrelevant in this regime, we conclude that the anomalously low value of $\alpha^*$ is a combined consequence of $\sigma$ band screening and excitonic shifts in the $\pi \rightarrow \pi^*$ continuum. 

\section{\label{sec:con}Conclusions}

We presented significantly improved measurements of the momentum- and energy- dependent effective fine structure constant of graphene, $\alpha^\ast(\ve{q},\omega)$, using inelastic x-ray scattering measurements of graphite. We deduce an asymptotic value of $\alpha^*$ in the range 0.25--0.35, which is larger than stated in our previous study \cite{reed} but smaller than the RPA value of 0.49, and implies an asymptotic value of $\epsilon$ in the range 6.6--8.8. We also performed $\pi$-band RPA calculations with and without interlayer hopping, and found that graphitic effects from interlayer hopping have no significant effect on screening in the range of momenta we studied. Screening by $\sigma$ band electrons, on the other hand, contribute significantly at finite wave vector. We therefore conclude that the anomalously low value of $\alpha^*$ may be attributed to a combination of excitonic effects in the $\pi \rightarrow \pi^*$ continuum and $\sigma$-band effects. The latter are very large at finite $q$, and should strongly screen short-ranged, Hubbard-like interactions in graphene and carbon allotropes more generally. 

We close by commenting on the discrepancies between our conclusions and the RPA calculation described in Ref. \cite{katsnelson}. That study reported a calculated static dielectric constant at $q = 0.238 \AA^{-1}$ that was not far from our earlier result derived from IXS experiments on graphite\cite{reed}. They attributed the anomalously low value of $\alpha^*$ to graphitic effects from the interlayer hopping. Here we have shown that the interlayer hopping parameters, $\gamma_1$ and $\gamma_3$, have no effect on the dielectric response at the lowest momenta measured in our experiments. The discrepancy between these two conclusions arises from the way the Coulomb interaction was handled in the two studies, which differ in two important ways. First, Ref. \cite{katsnelson} ignores the interaction between screening in the $\sigma$ and $\pi$ channels, which must be accounted for as described in our Appendix C. Second, in comparing to our result, they failed to eliminate the Coulomb interaction between the layers, which we have done through the use of Eq. 3. These differences result significantly larger values of $\epsilon(q,0)$ in Ref. \cite{katsnelson}, which they atribute to interlayer hopping effects. What we have shown is that, if the Coulomb effects are handled properly, the apparent effects of interlayer hopping are in fact small, and the primary causes of increased screening are excitonic effects and $\sigma$ band screening.

\begin{acknowledgments}
The authors acknowledge helpful discussions with L. K. Wagner. This work was supported by the U.S. Department of Energy grant DE-FG02-06ER46285, with use of the Advanced Photon Source supported by DEAC02-06CH11357. B.U. acknowledges NSF CAREER grant No. DMR-1352604 for support. E. F. acknowledges support from DOE award No. DE-SC001236. P.A. acknowledges support from the EPiQS Initiative of the Gordon and Betty Moore Foundation, through Grant GBMF4542.
\end{acknowledgments}

\appendix

\section{\label{app:sumfit}Data Processing}

Three steps of data processing are required to extract the response function, $\chi_{3D}(\ve{q},\omega)$, from the raw experimental spectra. The first is substraction of the quasielastic line, which is very intense in the experiment. The second is scaling the data using the $F$-sum rule, which provides the imaginary part, $\operatorname{Im}\chi_{3D}(\ve{q},\omega)$, in absolute units. The third is Kramers-Kronig analysis, which is needed to determine the real part, $\operatorname{Re}\chi_{3D}(\ve{q},\omega)$. 

\begin{figure*}
	\subfloat[][]{
		\includegraphics[width=0.45\textwidth]{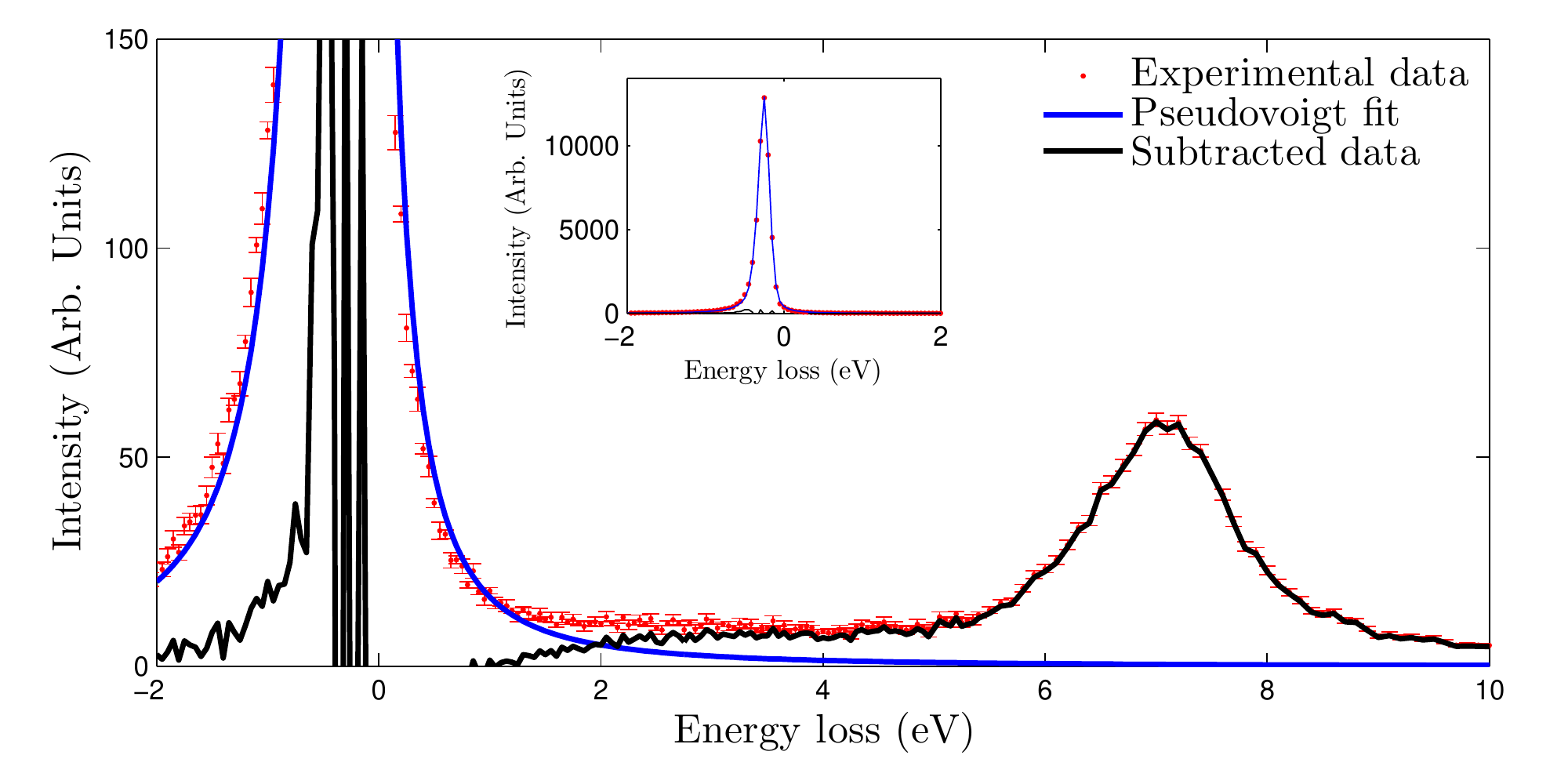}
		\label{fig:tth3subs}
		}\quad
	\subfloat[][]{
		\includegraphics[width=0.45\textwidth]{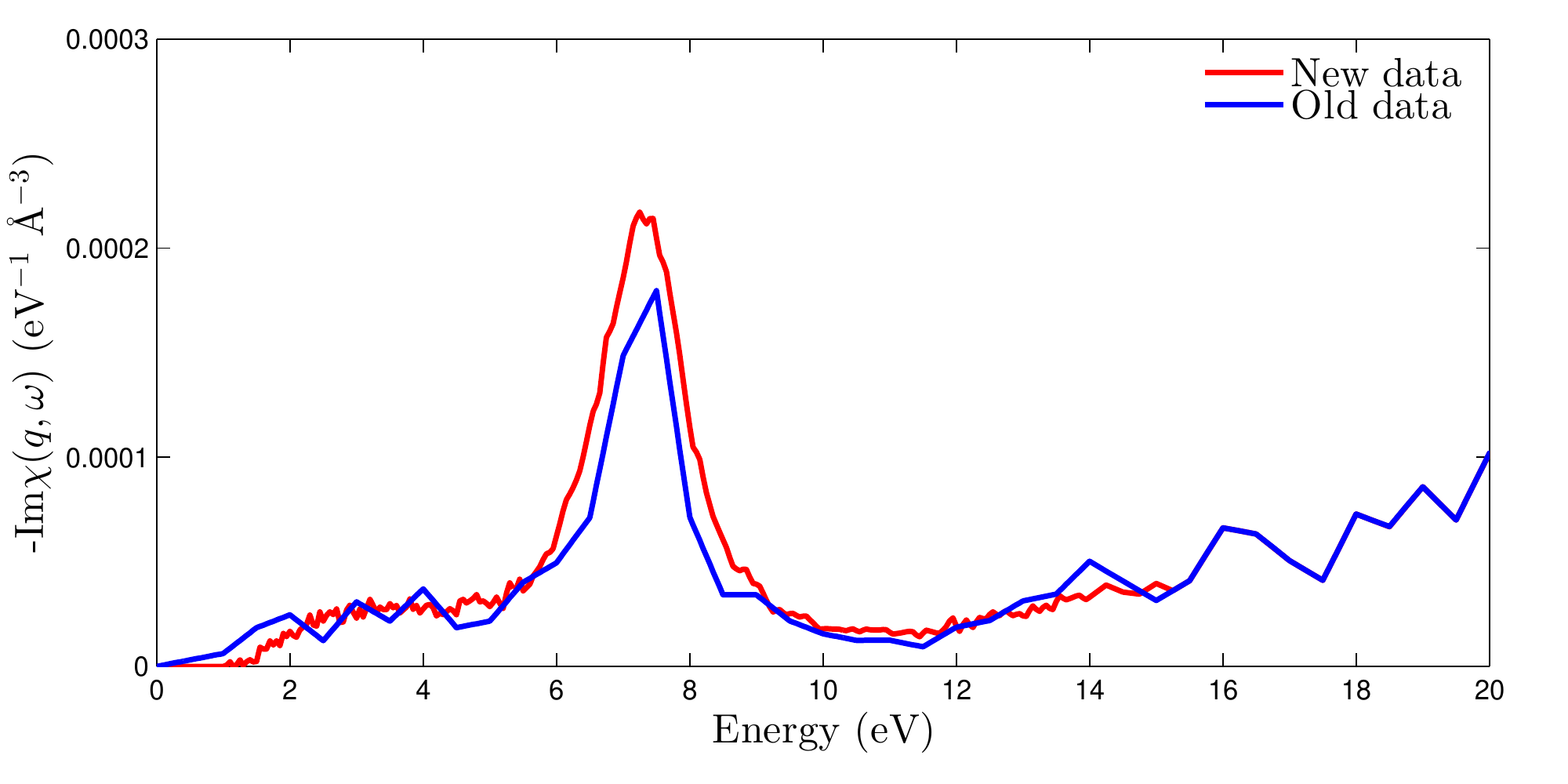}
		\label{fig:tth3norm}
		}\\
	\subfloat[][]{
		\includegraphics[width=0.45\textwidth]{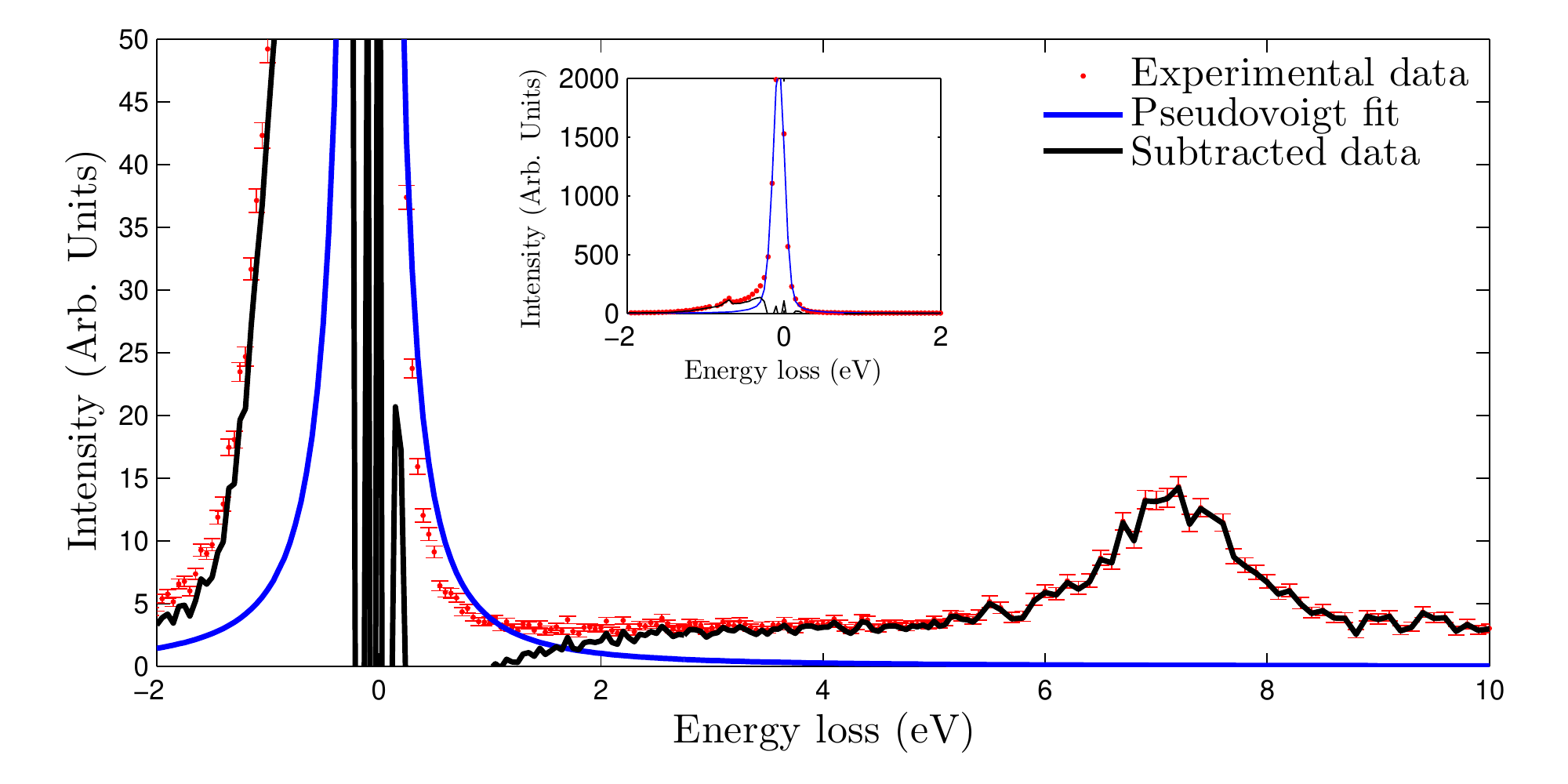}
		\label{fig:tth4subs}
		}\quad
	\subfloat[][]{
		\includegraphics[width=0.45\textwidth]{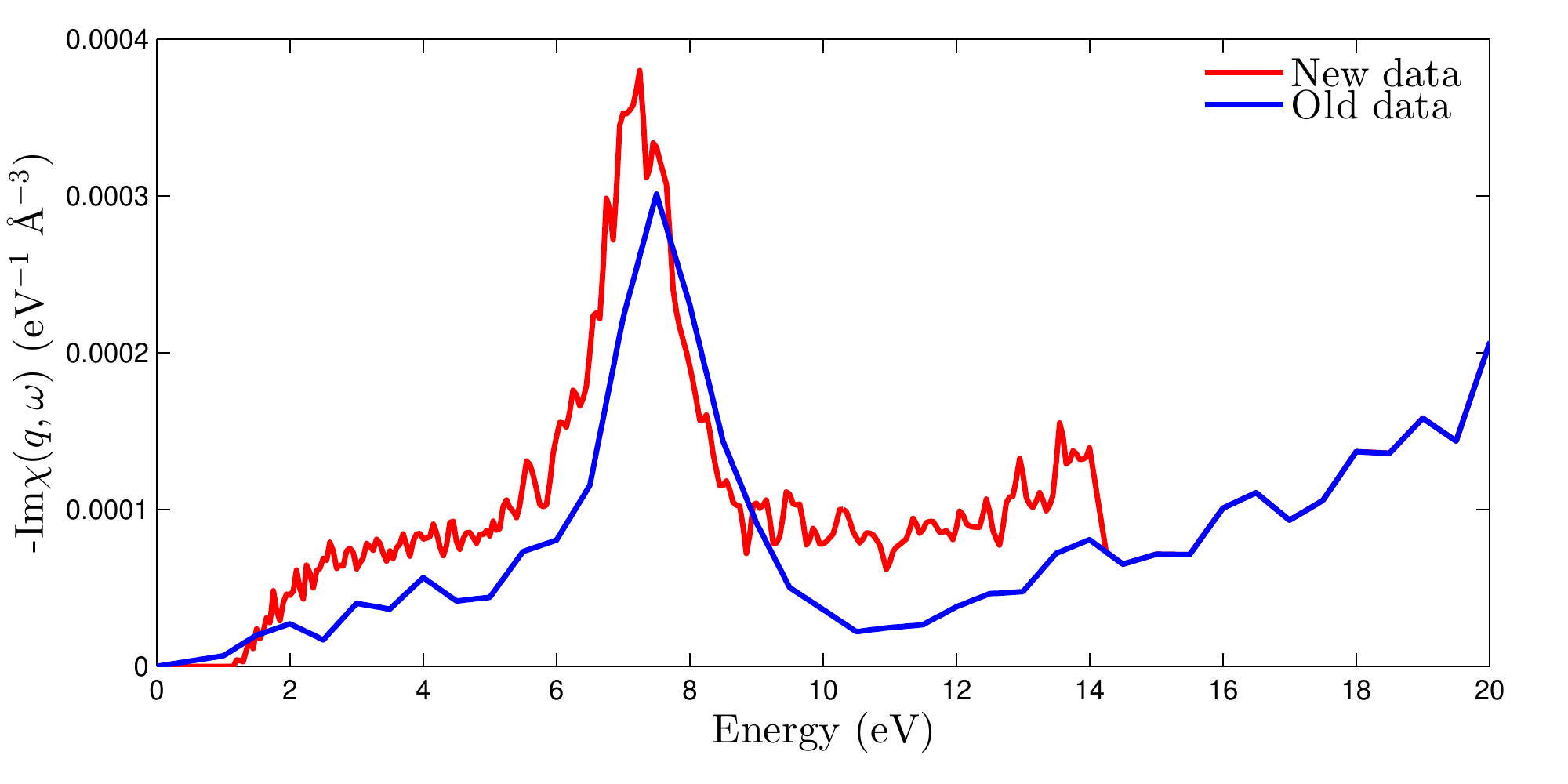}
		\label{fig:tth4norm}
		}
	\caption[]{Illustration of data processing steps described in Appendix A. (a) Subtraction of the elastic scattering from the IXS spectra. (red points) Raw IXS data for $q=0.212 \AA^{-1}$. (blue line) Fit to the elastic peak using a pseudo-Voigt function. (black points) Resulting spectrum, showing the onset of $\pi \rightarrow \pi^*$ excitations at $\sim$ 1 eV. (b) Matching of this spectrum to wide-range data from our previous study\cite{reed}, to aid in evaluting the sum rule integral. (c) Same illustration as in (a) for momentum $q=0.282 \AA^{-1}$. (d) Matching of the $q=0.282 \AA^{-1}$ spectrum to data from Ref. \cite{reed}, illustrating the mismatch present in a few of the spectra. This mismatch is incorporated into the determination of the error bars on the quantity $\operatorname{Im}\chi_{3D}(\ve{q},\omega)$.}
	\label{fig:specs}
\end{figure*}

To accomplish the first of the three, we used a pseudovoigt function---a linear combination of a Gaussian and a Lorentzian---to model the quasielastic lineshape:
\begin{equation}
	y = b+ A \times \left\{
	\begin{array}{lr} 
		c e^{-\frac{(x-x_0)^2}{2\sigma^2}}+ (1-c) \frac{l^2}{(x-x_0)^2 + l^2}, & x < x_0 \\
		c e^{-\frac{(x-x_0)^2}{2(\sigma \eta)^2}} + (1-c) \frac{(l \eta)^2}{(x-x_0)^2 +(l \eta)^2}, & x >= x_0.
	\end{array}
	\right.
	\label{eq:pvoigt}
\end{equation}
Here, $\sigma$ is the Gaussian width, $l = \sqrt{2\log 2}\sigma$ is the Lorentzian width, $c$ is a relative amplitude of the Guassian and Lorentzian components, $x_0$ is the center of the lineshape, and $b$ and $A$ are overall background and amplitude factors, respectively.  The parameter $\eta$ is an asymmetry factor that allows the positive- and negative-energy sides of the lineshape to be different, which is a feature in the experiment due to energy-dependent background scattering from the sample chamber.  In particular, the Lorentzian width $l$ is defined this way to best approximate a voigt function, which is a convolution of a Gaussian and Lorentzian.  In our experiments, background measurements on lithium fluoride, a large band gap material, showed $b=0$ for our setup, so this quantity was held fixed in our fits. Figs. \ref{fig:tth3subs} and \ref{fig:tth4subs} show subtractions for the two different momentum transfers $q = 0.212$ \AA$^{-1}$ and $q = 0.282$ \AA$^{-1}$, respectively.
To estimate the error introduced by subtracting the elastic line, we repeated the procedure with different constraints and initial parameters and recorded the range of values achieved in optimized fits. 

To normalize our data, we applied the $F$-sum rule over the entire momentum range. The results are shown in Fig. \ref{fig:imchi3d}. Energy-loss spectra from the current study were measured only over the interval $-4 < \omega < 15$ eV, so it was necessary to supplement using data from Ref. \cite{reed} to achieve a sufficiently wide range of energy to perform a Kramers-Kronig integration.  This was complicated by the fact that the $q$ values in the current study are not the same as those in from Ref. \cite{reed}.  To overcome this mismatch, we linearly interpolated the data from Ref. \cite{reed} to match our $q$ values.  To properly normalize and patch our spectra to the previous spectra, we first note that in Ref. \cite{reed}, the proportionality between the measured spectra and $S(\ve{q},\omega)$ (and, thus, $\chi(\ve{q},\omega)$) was determined by fitting the spectra to the $F$-sum rule.  In our new experiment, for each separate experimental run, we took our spectra, with the exception of the $q=0.212$ \AA$^{-1}$, and fit to the spectra of Ref. \cite{reed}, finding a multiplicative factor for each momentum transfer. We then used the average of these multiplicative factors to scale the complete data set.  This was done because the multiplicative factor is determined predominantly by incident beam features including intensity, angular divergence, and longitudinal bandwidth, as well as sample thickness, which are features that are constant or nearly constant within a single experimental run, but may vary from one beam time to the next.  For $q=0.212$ \AA$^{-1}$ we took the average multiplicative factor for the $q=0.352$ \AA$^{-1}$ and $q=0.563$ \AA$^{-1}$ spectra (the other two momentum transfers from that run) and defined that its multiplicative factor rather than directly fitting to the old data; this is because the lowest momentum value from the old data was $q=0.238$ \AA$^{-1}$, and extrapolation to lower momenta is not reliable.  Figs. \ref{fig:tth3norm} and \ref{fig:tth4norm} show our final, normalized spectra for $q = 0.212$ \AA$^{-1}$ and $q = 0.282$ \AA$^{-1}$, respectively, in absolute units. After acquiring these spectra, Kramers-Kronig analysis was straight forward using standard methods. 

The main advantage of our data processing procedure is that it fits the $\pi$-plasmon intensities well, which is ideal given that $\pi$-plasmon is the strongest contribution to the Kramers-Kronig integral in the energy range we measure.  Unfortunately, for a few momentum values, our method introduced discontinuities at the patch energy ($\approx$ 15 eV), as illustrated in Fig. \ref{fig:tth4norm}. There being no objective way to correct for this, we simply included the effect in our estimate of the error bars on the value of $\chi(\ve{q},\omega)$ (see Fig. 1a).  Indeed, the spectra in Fig. \ref{fig:chiplots} represent the average values of $\chi$ after elastic line subtraction and intensity normalization, and the error bars are a sum in quadrature of the maximum positive or negative deviation from the mean and a 5\% estimate of the variance of the sum-rule normalization.

\section{\label{app:rpa}RPA}

As discussed in the main text, RPA calculations for graphite and graphene are based on a tight-binding model of the $\pi$ electrons. The graphite calculations assumed ABA(Bernal) stacking, illustrated in Fig. \ref{fig:structure}, including nearest-neighbor interlayer hopping. The tight-binding Hamiltonian has the explicit form
\begin{align}
	H =
		 & -t \displaystyle\sum^{N_{\textrm{layer}}}_{l,\braket{i,j}}
		\left[a^{\dagger}_{i,l}b_{j,l}+b^{\dagger}_{j,l}a_{i,l}\right] \nonumber \\
		 & -\gamma_1 \left[\displaystyle\sum_{l_{\textrm{even}},\braket{\braket{j,j'}}}
		\left[a^{\dagger}_{j,l}b_{j',l+1}+b^{\dagger}_{j',l+1}a_{j,l}\right]\right.\nonumber\\
		& \left.+\displaystyle\sum_{l_{\mro},\braket{\braket{j,j'}}}
		\left[b^{\dagger}_{j,l}a_{j',l+1}+a^{\dagger}_{j',l+1}b_{j,l}\right]
		\right] \nonumber \\
		 & -\gamma_3 \left[\displaystyle\sum_{l_{\textrm{even}},\braket{\braket{\braket{j,j'}}}}
		\left[b^{\dagger}_{j,l}a_{j',l+1}+a^{\dagger}_{j',l+1}b_{j,l}\right]
		\right. \nonumber \\
		& \left.+ \displaystyle\sum_{l_{\mro},\braket{\braket{\braket{j,j'}}}}
		\left[a^{\dagger}_{j,l}b_{j',l+1}+b^{\dagger}_{j',l+1}a_{j,l}\right]
\right]
\label{eq:hgraphite}
\end{align}
where the sum is taken over nearest neighbors $\braket{i,j}$ and the layers, $l$, $t=3$ eV is the in-plane hopping parameter, $\gamma_1=0.4$ eV is the vertical hopping between the $A$ and $B$ sublattice (indicated by the double brackets), and $\gamma_3=0.3$ eV is the nearest $A$ and $B$ interlayer coupling (indicated by the triple brackets) (see Fig. 2a).  Single layer graphene can be described by the first sum in Eq. \ref{eq:hgraphite}.  For graphene, the wavefunctions can be computed explicitly, and we can write the overlap function as 
\begin{equation}
	f_{s\cdot s'}(\ve{k},\ve{q}_{\parallel}) = \left | \frac{\phi^*(\ve{k})}{|\phi(\ve{k})|} 
		\frac{\phi(\ve{k}+\ve{q})}{|\phi(\ve{k}+\ve{q})} + s\cdot s' \right |^2,
	\label{eq:goverlap}
\end{equation}
where
\begin{equation}
	\phi(\ve{k}) = e^{-ik_x a} + 2e^{-i k_x a / 2}\cos\left(\frac{\sqrt{3}}{2}k_y a \right),
	\label{eq:phi}
\end{equation}
and $a=1.42 \AA$ is the in-plane carbon-carbon distance.  

Once the tight-binding wave functions were obtained, the polarization functions were obtained by numerically evaluating the Lindhard formula, 
\begin{align}
	\displaystyle
	\Pi&(\ve{q}_{\parallel},\omega) = \frac{g_s}{(2\pi)^d}\int_{BZ} d^2\ve{k} \nonumber\\
	&\times
	\sum_{s,s' = \pm 1} f_{s\cdot s'}(\ve{k},\ve{q}_{\parallel})\frac{n_F[E^s(\ve{k})] - n_F[E^{s'}(\ve{k}+\ve{q})]}
	{E^s(\ve{k}) - E^{s'}(\ve{k}+\ve{q}) + \omega + i\eta}.
\label{eq:pirpa}
\end{align}
where the integral is over the Brillouin zone, $d$ is the dimensionality (either 2 or 3 for graphene or graphite, respectively), $g_s$ is the spin degeneracy, $E^{s/s'}(\ve{k})$ is the energy dispersion for the conduction ($s = 1$) and valence ($s=-1$) $\pi$-bands, $\eta$ is a convergence factor, and $f_{s\cdot s'}(\ve{k},\ve{q}_{\parallel})$ describes the overlap between the electron and hole wavefunctions (Eq. \ref{eq:goverlap}. We use the full $\pi$-band wavefunctions (as opposed to linearized bands\cite{katsnelson}) to compute the responses of both graphene and graphite. 
All of our calculations use $\theta = 0^{\circ}$ (Fig. \ref{fig:brillouin}), since the experimental data was found to be insensitive to this angle at the low momenum values studied.  We performed calculations for both $30^{\circ}$ and $60^{\circ}$ directions as well, and the differences were found to be within numerical uncertainty. The results, shown in Fig. 3 for a convergence factor of $\eta = 0.01$ eV, agree closely with those of Ref. \cite{katsnelson}.
 
For graphite, there are four symmetry-inequivalent sublattices, A and B, leading to four basis functions rather than the two of graphene.  There is no simple analytic expression for these wavefunctions, so to perform the integral in Eq. \ref{eq:pirpa} we simply find the eigenfunctions of the graphite Hamiltonian, and use those to compute the corresponding energies and overlap functions $f_{s\cdot s'}$.  It is then straightforward to numerically integrate Eq. \ref{eq:pirpa} over the 3D Brillouin zone of graphite to obtain $\Pi_{3D}(\ve{q},\omega)$.

\section{\label{app:kappa} Dielectric function}

It is important in our analysis to properly describe the interaction between screening processes involving $\pi$ electrons and $\sigma$ electrons. 
To do so, we treat the electrons from the $\pi$-bands and $\sigma$-bands as
two separate fermionic species. In general, we can write the Hamiltonian
for such a system as 
\begin{equation}
H=H_{0}+H_{I}+H_{U},\label{eq:htrace}
\end{equation}
where 
\begin{equation}
{\displaystyle H_{0}=\sum_{\ve{k}\sigma=\uparrow\downarrow}E_{\ve{k}}c_{\ve{k}\sigma}^{\dagger}c_{\ve{k}\sigma}+\sum_{\ve{k}\sigma=\uparrow\downarrow}\epsilon_{\ve{k}}g_{\ve{k}\sigma}^{\dagger}g_{\ve{k}\sigma}\label{eq:htrace0}}
\end{equation}
describes the noninteracting Hamiltonian for the fermionic operators
of the $\pi$- and $\sigma$-electrons, $g(g^{\dagger})$ and $c(c^{\dagger})$,
respectively, with $\epsilon_{\ve{k}}$ and $E_{\ve{k}}$ their corresponding
energy spectra. 
\begin{align}
{\displaystyle H_{I}=} & \frac{1}{2}\sum_{\ve{q}}V(\ve{q})\hat{\rho}_{\pi}(-\ve{q})\hat{\rho}_{\pi}(\ve{q})+\sum_{\ve{q}}V(\ve{q})\hat{\rho}_{\pi}(-\ve{q})\hat{\rho}_{\sigma}(\ve{q})\nonumber \\
 & +\frac{1}{2}\sum_{\ve{q}}V(\ve{q})\hat{\rho}_{\sigma}(-\ve{q})\hat{\rho}_{\sigma}(\ve{q}),\label{eq:htracei}
\end{align}
is the Coulomb interaction energy, with $\hat{\rho}_{\pi}$ and $\hat{\rho}_{\sigma}$
the corresponding particle density operators of the $\pi$ and $\sigma$
electrons, $V(q)$ is the Coulomb interaction (see main text), and
\begin{equation}
{\displaystyle H_{U}=\sum_{\ve{q}}U(\ve{q})\left[\hat{\rho}_{\sigma}(\ve{q})+\hat{\rho}_{\pi}(\ve{q})\right]\label{eq:htraceu}}
\end{equation}
is an external potential that couples to the density. 

At the RPA level, the equations of motion for the density of $\pi$
and $\sigma$ electrons
\begin{equation}
i\frac{\textrm{d}\langle\hat{\rho}_{\pi}(\mathbf{q})\rangle}{\textrm{d}t}=\langle[H,\hat{\rho}_{\pi}(\mathbf{q})]\rangle\label{eq:eqmotionrho}
\end{equation}
\begin{equation}
i\frac{\textrm{d}\langle\hat{\rho}_{\sigma}(\mathbf{q})\rangle}{\textrm{d}t}=\langle[H,\hat{\rho}_{\sigma}(\mathbf{q})]\rangle\label{eq:eqmotionrhos}
\end{equation}
 can be calculated in linear response theory \cite{pines} and written
in matrix form as 
\begin{equation}
\left(\begin{array}{c}
\braket{\hat{\rho}_{\sigma}(\ve{q},\omega)}\\
\braket{\hat{\rho}_{\pi}(\ve{q},\omega)}
\end{array}\right)=\frac{U(\ve{q})}{\epsilon(\ve{q},\omega)}\mathbf{M}\left(\begin{array}{c}
\Pi_{\sigma}(\ve{q},\omega)\\
\Pi_{\pi}(\ve{q},\omega),
\end{array}\right)\label{eq:htracemat}
\end{equation}
where 
\begin{equation}
\mathbf{M}=\left(\begin{array}{cc}
1-V(\ve{q})\Pi_{\pi}(\ve{q},\omega) & V(\ve{q})\Pi_{\sigma}(\ve{q},\omega)\\
V(\ve{q})\Pi_{\pi}(\ve{q},\omega) & 1-V(\ve{q})\Pi_{\sigma}(\ve{q},\omega)
\end{array}\right)\label{eq:htracem}
\end{equation}
and 
\begin{align*}
\epsilon(\ve{q},\omega) & =\operatorname{det}\mathbf{M}\\
 & =1-V(\ve{q})\Pi_{\pi}(\ve{q},\omega)-V(\ve{q})\Pi_{\sigma}(\ve{q},\omega)
\end{align*}
is the dielectric function of the system, where $\Pi_{\pi}$ and $\Pi_{s}$
are the RPA polarization functions of the $\pi$- and $\sigma$-electrons,
respectively. Equation (\ref{eq:htracemat}) is completely general
and is valid for any two band system. The dielectric function can
be equivalently written as 
\[
\epsilon(\ve{q},\omega)=\kappa_{\sigma}(\ve{q})-V(\ve{q})\Pi_{\pi}(\ve{q},\omega),
\]
where we have defined 
\begin{equation}
\kappa_{\sigma}(\ve{q})=1-V(\ve{q})\Pi_{\sigma}(\ve{q},\omega)
\end{equation}
as the background dielectric constant due to the $\sigma$-electrons.
In this way we recover Eq. \ref{eq:epsilon} for the total dielectric
function of graphene and graphite.

Eq. \ref{eq:htracemat} simplifies in two equations:
\begin{equation}
\braket{\hat{\rho}_{\sigma}(\ve{q},i\omega)}=U(\ve{q})\frac{\Pi_{\sigma}(\ve{q},\omega)}{\epsilon(\ve{q},\omega)}\label{eq:rhos}
\end{equation}
and 
\begin{equation}
\braket{\hat{\rho}_{\pi}(\ve{q},i\omega)}=U(\ve{q})\frac{\Pi_{\pi}(\ve{q},\omega)}{\epsilon(\ve{q},\omega)}.\label{eq:rhopi}
\end{equation}
The first equation defines the susceptibility of the sigma electrons
in the presence of screening effects from both bands,
\begin{equation}
\chi_{\sigma}(\ve{q},\omega)=\frac{\Pi_{\sigma}(\ve{q},\omega)}{\epsilon(\ve{q},i\omega)}.\label{eq:chis}
\end{equation}
The second one gives the susceptibility of the $\pi$ electrons in
the presence of the sigma bands:
\begin{equation}
\chi_{\pi}(\ve{q},\omega)=\frac{\Pi_{\pi}(\ve{q},\omega)}{\epsilon(\ve{q},\omega)}.\label{eq:chipi-1}
\end{equation}
 The full susceptibility is defined by the response function including
both $\sigma$ and $\pi$ electrons density fluctuations, 
\begin{equation}
\braket{\hat{\rho}_{\pi}(\ve{q},i\omega)}+\braket{\hat{\rho}_{\sigma}(\ve{q},i\omega)}=U(\ve{q})\chi(\ve{q},\omega),\label{eq:rhoT}
\end{equation}
where 
\begin{equation}
\chi(\ve{q},\omega)=\chi_{\pi}(\ve{q},\omega)+\chi_{\sigma}(\ve{q},\omega)=\frac{\Pi_{\pi}(\ve{q},\omega)+\Pi_{\sigma}(\ve{q},\omega)}{\epsilon(\ve{q},\omega)}.\label{eq:chi}
\end{equation}

\bibliography{graphpaper}

\begin{thebibliography}{25}%
\makeatletter
\providecommand \@ifxundefined [1]{%
 \@ifx{#1\undefined}
}%
\providecommand \@ifnum [1]{%
 \ifnum #1\expandafter \@firstoftwo
 \else \expandafter \@secondoftwo
 \fi
}%
\providecommand \@ifx [1]{%
 \ifx #1\expandafter \@firstoftwo
 \else \expandafter \@secondoftwo
 \fi
}%
\providecommand \natexlab [1]{#1}%
\providecommand \enquote  [1]{``#1''}%
\providecommand \bibnamefont  [1]{#1}%
\providecommand \bibfnamefont [1]{#1}%
\providecommand \citenamefont [1]{#1}%
\providecommand \href@noop [0]{\@secondoftwo}%
\providecommand \href [0]{\begingroup \@sanitize@url \@href}%
\providecommand \@href[1]{\@@startlink{#1}\@@href}%
\providecommand \@@href[1]{\endgroup#1\@@endlink}%
\providecommand \@sanitize@url [0]{\catcode `\\12\catcode `\$12\catcode
  `\&12\catcode `\#12\catcode `\^12\catcode `\_12\catcode `\%12\relax}%
\providecommand \@@startlink[1]{}%
\providecommand \@@endlink[0]{}%
\providecommand \url  [0]{\begingroup\@sanitize@url \@url }%
\providecommand \@url [1]{\endgroup\@href {#1}{\urlprefix }}%
\providecommand \urlprefix  [0]{URL }%
\providecommand \Eprint [0]{\href }%
\providecommand \doibase [0]{http://dx.doi.org/}%
\providecommand \selectlanguage [0]{\@gobble}%
\providecommand \bibinfo  [0]{\@secondoftwo}%
\providecommand \bibfield  [0]{\@secondoftwo}%
\providecommand \translation [1]{[#1]}%
\providecommand \BibitemOpen [0]{}%
\providecommand \bibitemStop [0]{}%
\providecommand \bibitemNoStop [0]{.\EOS\space}%
\providecommand \EOS [0]{\spacefactor3000\relax}%
\providecommand \BibitemShut  [1]{\csname bibitem#1\endcsname}%
\let\auto@bib@innerbib\@empty
\bibitem [{\citenamefont {Reed}\ \emph {et~al.}(2010)\citenamefont {Reed} \emph
  {et~al.}}]{reed}%
  \BibitemOpen
  \bibfield  {author} {\bibinfo {author} {\bibfnamefont {J.~P.}\ \bibnamefont
  {Reed}} \emph {et~al.},\ }\href {\doibase 10.1126/science.1190920} {\bibfield
   {journal} {\bibinfo  {journal} {Science}\ }\textbf {\bibinfo {volume}
  {330}},\ \bibinfo {pages} {805} (\bibinfo {year} {2010})}\BibitemShut
  {NoStop}%
\bibitem [{\citenamefont {Castro~Neto}\ \emph {et~al.}(2009)\citenamefont
  {Castro~Neto}, \citenamefont {Guinea}, \citenamefont {Peres}, \citenamefont
  {Novoselov},\ and\ \citenamefont {Geim}}]{graphenereview}%
  \BibitemOpen
  \bibfield  {author} {\bibinfo {author} {\bibfnamefont {A.~H.}\ \bibnamefont
  {Castro~Neto}}, \bibinfo {author} {\bibfnamefont {F.}~\bibnamefont {Guinea}},
  \bibinfo {author} {\bibfnamefont {N.~M.~R.}\ \bibnamefont {Peres}}, \bibinfo
  {author} {\bibfnamefont {K.~S.}\ \bibnamefont {Novoselov}}, \ and\ \bibinfo
  {author} {\bibfnamefont {A.~K.}\ \bibnamefont {Geim}},\ }\href {\doibase
  10.1103/RevModPhys.81.109} {\bibfield  {journal} {\bibinfo  {journal} {Rev.
  Mod. Phys.}\ }\textbf {\bibinfo {volume} {81}},\ \bibinfo {pages} {109}
  (\bibinfo {year} {2009})}\BibitemShut {NoStop}%
\bibitem [{\citenamefont {Das~Sarma}\ \emph {et~al.}(2011)\citenamefont
  {Das~Sarma}, \citenamefont {Adam}, \citenamefont {Hwang},\ and\ \citenamefont
  {Rossi}}]{transportrev}%
  \BibitemOpen
  \bibfield  {author} {\bibinfo {author} {\bibfnamefont {S.}~\bibnamefont
  {Das~Sarma}}, \bibinfo {author} {\bibfnamefont {S.}~\bibnamefont {Adam}},
  \bibinfo {author} {\bibfnamefont {E.~H.}\ \bibnamefont {Hwang}}, \ and\
  \bibinfo {author} {\bibfnamefont {E.}~\bibnamefont {Rossi}},\ }\href
  {\doibase 10.1103/RevModPhys.83.407} {\bibfield  {journal} {\bibinfo
  {journal} {Rev. Mod. Phys.}\ }\textbf {\bibinfo {volume} {83}},\ \bibinfo
  {pages} {407} (\bibinfo {year} {2011})}\BibitemShut {NoStop}%
\bibitem [{\citenamefont {Wallace}(1947)}]{wallace}%
  \BibitemOpen
  \bibfield  {author} {\bibinfo {author} {\bibfnamefont {P.~R.}\ \bibnamefont
  {Wallace}},\ }\href {\doibase 10.1103/PhysRev.71.622} {\bibfield  {journal}
  {\bibinfo  {journal} {Phys. Rev.}\ }\textbf {\bibinfo {volume} {71}},\
  \bibinfo {pages} {622} (\bibinfo {year} {1947})}\BibitemShut {NoStop}%
\bibitem [{\citenamefont {Semenoff}(1984)}]{dirac}%
  \BibitemOpen
  \bibfield  {author} {\bibinfo {author} {\bibfnamefont {G.~W.}\ \bibnamefont
  {Semenoff}},\ }\href {\doibase 10.1103/PhysRevLett.53.2449} {\bibfield
  {journal} {\bibinfo  {journal} {Phys. Rev. Lett.}\ }\textbf {\bibinfo
  {volume} {53}},\ \bibinfo {pages} {2449} (\bibinfo {year}
  {1984})}\BibitemShut {NoStop}%
\bibitem [{\citenamefont {Kotov}\ \emph {et~al.}(2012)\citenamefont {Kotov},
  \citenamefont {Uchoa}, \citenamefont {Pereira}, \citenamefont {Guinea},\ and\
  \citenamefont {Neto}}]{kotov}%
  \BibitemOpen
  \bibfield  {author} {\bibinfo {author} {\bibfnamefont {V.~N.}\ \bibnamefont
  {Kotov}}, \bibinfo {author} {\bibfnamefont {B.}~\bibnamefont {Uchoa}},
  \bibinfo {author} {\bibfnamefont {V.~M.}\ \bibnamefont {Pereira}}, \bibinfo
  {author} {\bibfnamefont {F.}~\bibnamefont {Guinea}}, \ and\ \bibinfo {author}
  {\bibfnamefont {A.~H.~C.}\ \bibnamefont {Neto}},\ }\href {\doibase
  10.1103/RevModPhys.84.1067} {\bibfield  {journal} {\bibinfo  {journal} {Rev.
  Mod. Phys.}\ }\textbf {\bibinfo {volume} {84}},\ \bibinfo {pages} {1067}
  (\bibinfo {year} {2012})}\BibitemShut {NoStop}%
\bibitem [{\citenamefont {Ulybyshev}\ \emph {et~al.}(2013)\citenamefont
  {Ulybyshev}, \citenamefont {Buividovich}, \citenamefont {Katsnelson},\ and\
  \citenamefont {Polikarpov}}]{ulybyshev}%
  \BibitemOpen
  \bibfield  {author} {\bibinfo {author} {\bibfnamefont {M.~V.}\ \bibnamefont
  {Ulybyshev}}, \bibinfo {author} {\bibfnamefont {P.~V.}\ \bibnamefont
  {Buividovich}}, \bibinfo {author} {\bibfnamefont {M.~I.}\ \bibnamefont
  {Katsnelson}}, \ and\ \bibinfo {author} {\bibfnamefont {M.~I.}\ \bibnamefont
  {Polikarpov}},\ }\href@noop {} {\bibfield  {journal} {\bibinfo  {journal}
  {Phys. Rev. Lett.}\ }\textbf {\bibinfo {volume} {111}},\ \bibinfo {pages}
  {056801} (\bibinfo {year} {2013})}\BibitemShut {NoStop}%
\bibitem [{\citenamefont {Sodemann}\ and\ \citenamefont
  {Fogler}(2012)}]{fogler}%
  \BibitemOpen
  \bibfield  {author} {\bibinfo {author} {\bibfnamefont {I.}~\bibnamefont
  {Sodemann}}\ and\ \bibinfo {author} {\bibfnamefont {M.~M.}\ \bibnamefont
  {Fogler}},\ }\href {\doibase 10.1103/PhysRevB.86.115408} {\bibfield
  {journal} {\bibinfo  {journal} {Phys. Rev. B}\ }\textbf {\bibinfo {volume}
  {86}},\ \bibinfo {pages} {115408} (\bibinfo {year} {2012})}\BibitemShut
  {NoStop}%
\bibitem [{\citenamefont {Gonzalez}\ \emph {et~al.}(1994)\citenamefont
  {Gonzalez}, \citenamefont {Guinea},\ and\ \citenamefont
  {Vozmediano}}]{non-fermi_renorm}%
  \BibitemOpen
  \bibfield  {author} {\bibinfo {author} {\bibfnamefont {J.}~\bibnamefont
  {Gonzalez}}, \bibinfo {author} {\bibfnamefont {F.}~\bibnamefont {Guinea}}, \
  and\ \bibinfo {author} {\bibfnamefont {M.}~\bibnamefont {Vozmediano}},\
  }\href {\doibase http://dx.doi.org/10.1016/0550-3213(94)90410-3} {\bibfield
  {journal} {\bibinfo  {journal} {Nuclear Physics B}\ }\textbf {\bibinfo
  {volume} {424}},\ \bibinfo {pages} {595 } (\bibinfo {year}
  {1994})}\BibitemShut {NoStop}%
\bibitem [{\citenamefont {Mishchenko}(2007)}]{mishchenko}%
  \BibitemOpen
  \bibfield  {author} {\bibinfo {author} {\bibfnamefont {E.~G.}\ \bibnamefont
  {Mishchenko}},\ }\href@noop {} {\bibfield  {journal} {\bibinfo  {journal}
  {Phys. Rev. Lett.}\ }\textbf {\bibinfo {volume} {98}},\ \bibinfo {pages}
  {216801} (\bibinfo {year} {2007})}\BibitemShut {NoStop}%
\bibitem [{\citenamefont {Hasan}\ and\ \citenamefont {Kane}(2010)}]{kane}%
  \BibitemOpen
  \bibfield  {author} {\bibinfo {author} {\bibfnamefont {M.~Z.}\ \bibnamefont
  {Hasan}}\ and\ \bibinfo {author} {\bibfnamefont {C.~L.}\ \bibnamefont
  {Kane}},\ }\href@noop {} {\bibfield  {journal} {\bibinfo  {journal} {Rev.
  Mod. Phys.}\ }\textbf {\bibinfo {volume} {82}},\ \bibinfo {pages} {3045}
  (\bibinfo {year} {2010})}\BibitemShut {NoStop}%
\bibitem [{\citenamefont {Qi}\ and\ \citenamefont {Zhang}(2011)}]{qi}%
  \BibitemOpen
  \bibfield  {author} {\bibinfo {author} {\bibfnamefont {X.-L.}\ \bibnamefont
  {Qi}}\ and\ \bibinfo {author} {\bibfnamefont {S.-C.}\ \bibnamefont {Zhang}},\
  }\href@noop {} {\bibfield  {journal} {\bibinfo  {journal} {Rev. Mod. Phys.}\
  }\textbf {\bibinfo {volume} {83}},\ \bibinfo {pages} {1057} (\bibinfo {year}
  {2011})}\BibitemShut {NoStop}%
\bibitem [{\citenamefont {Rossnagel}(2011)}]{rossnagel}%
  \BibitemOpen
  \bibfield  {author} {\bibinfo {author} {\bibfnamefont {K.}~\bibnamefont
  {Rossnagel}},\ }\href@noop {} {\bibfield  {journal} {\bibinfo  {journal} {J.
  Phys.: Condens. Matter}\ }\textbf {\bibinfo {volume} {23}},\ \bibinfo {pages}
  {213001} (\bibinfo {year} {2011})}\BibitemShut {NoStop}%
\bibitem [{\citenamefont {Wan}\ \emph {et~al.}(2011)\citenamefont {Wan},
  \citenamefont {Turner}, \citenamefont {Vishwanath},\ and\ \citenamefont
  {Savrasov}}]{ashvin}%
  \BibitemOpen
  \bibfield  {author} {\bibinfo {author} {\bibfnamefont {X.}~\bibnamefont
  {Wan}}, \bibinfo {author} {\bibfnamefont {A.~M.}\ \bibnamefont {Turner}},
  \bibinfo {author} {\bibfnamefont {A.}~\bibnamefont {Vishwanath}}, \ and\
  \bibinfo {author} {\bibfnamefont {S.~Y.}\ \bibnamefont {Savrasov}},\
  }\href@noop {} {\bibfield  {journal} {\bibinfo  {journal} {Phys. Rev. B}\
  }\textbf {\bibinfo {volume} {83}},\ \bibinfo {pages} {205101} (\bibinfo
  {year} {2011})}\BibitemShut {NoStop}%
\bibitem [{\citenamefont {Young}\ \emph {et~al.}(2012)\citenamefont {Young},
  \citenamefont {Zaheer}, \citenamefont {Teo}, \citenamefont {Kane},
  \citenamefont {Mele},\ and\ \citenamefont {Rappe}}]{young}%
  \BibitemOpen
  \bibfield  {author} {\bibinfo {author} {\bibfnamefont {S.~M.}\ \bibnamefont
  {Young}}, \bibinfo {author} {\bibfnamefont {S.}~\bibnamefont {Zaheer}},
  \bibinfo {author} {\bibfnamefont {J.~C.~Y.}\ \bibnamefont {Teo}}, \bibinfo
  {author} {\bibfnamefont {C.~L.}\ \bibnamefont {Kane}}, \bibinfo {author}
  {\bibfnamefont {E.~J.}\ \bibnamefont {Mele}}, \ and\ \bibinfo {author}
  {\bibfnamefont {A.~M.}\ \bibnamefont {Rappe}},\ }\href@noop {} {\bibfield
  {journal} {\bibinfo  {journal} {Phys. Rev. Lett.}\ }\textbf {\bibinfo
  {volume} {108}},\ \bibinfo {pages} {140405} (\bibinfo {year}
  {2012})}\BibitemShut {NoStop}%
\bibitem [{\citenamefont {Yuan}\ \emph {et~al.}(2011)\citenamefont {Yuan},
  \citenamefont {Rold\'an},\ and\ \citenamefont {Katsnelson}}]{katsnelson}%
  \BibitemOpen
  \bibfield  {author} {\bibinfo {author} {\bibfnamefont {S.}~\bibnamefont
  {Yuan}}, \bibinfo {author} {\bibfnamefont {R.}~\bibnamefont {Rold\'an}}, \
  and\ \bibinfo {author} {\bibfnamefont {M.~I.}\ \bibnamefont {Katsnelson}},\
  }\href {\doibase 10.1103/PhysRevB.84.035439} {\bibfield  {journal} {\bibinfo
  {journal} {Phys. Rev. B}\ }\textbf {\bibinfo {volume} {84}},\ \bibinfo
  {pages} {035439} (\bibinfo {year} {2011})}\BibitemShut {NoStop}%
\bibitem [{\citenamefont {Shung}(1986)}]{shung}%
  \BibitemOpen
  \bibfield  {author} {\bibinfo {author} {\bibfnamefont {K.~W.-K.}\
  \bibnamefont {Shung}},\ }\href@noop {} {\bibfield  {journal} {\bibinfo
  {journal} {Phys. Rev. B}\ }\textbf {\bibinfo {volume} {34}},\ \bibinfo
  {pages} {979} (\bibinfo {year} {1986})}\BibitemShut {NoStop}%
\bibitem [{\citenamefont {Mak}\ \emph {et~al.}(2011)\citenamefont {Mak},
  \citenamefont {Shan},\ and\ \citenamefont {Heinz}}]{makHeinzExc}%
  \BibitemOpen
  \bibfield  {author} {\bibinfo {author} {\bibfnamefont {K.~F.}\ \bibnamefont
  {Mak}}, \bibinfo {author} {\bibfnamefont {J.}~\bibnamefont {Shan}}, \ and\
  \bibinfo {author} {\bibfnamefont {T.~F.}\ \bibnamefont {Heinz}},\ }\href@noop
  {} {\bibfield  {journal} {\bibinfo  {journal} {Phys. Rev. Lett.}\ }\textbf
  {\bibinfo {volume} {106}},\ \bibinfo {pages} {046401} (\bibinfo {year}
  {2011})}\BibitemShut {NoStop}%
\bibitem [{\citenamefont {Wehling}\ \emph {et~al.}(2011)\citenamefont
  {Wehling}, \citenamefont {\ifmmode \mbox{\c{S}}\else \c{S}\fi{}a\ifmmode
  \mbox{\c{s}}\else \c{s}\fi{}\ifmmode \imath \else \i
  \fi{}o\ifmmode~\breve{g}\else \u{g}\fi{}lu}, \citenamefont {Friedrich},
  \citenamefont {Lichtenstein}, \citenamefont {Katsnelson},\ and\ \citenamefont
  {Bl\"ugel}}]{kappakats}%
  \BibitemOpen
  \bibfield  {author} {\bibinfo {author} {\bibfnamefont {T.~O.}\ \bibnamefont
  {Wehling}}, \bibinfo {author} {\bibfnamefont {E.}~\bibnamefont {\ifmmode
  \mbox{\c{S}}\else \c{S}\fi{}a\ifmmode \mbox{\c{s}}\else \c{s}\fi{}\ifmmode
  \imath \else \i \fi{}o\ifmmode~\breve{g}\else \u{g}\fi{}lu}}, \bibinfo
  {author} {\bibfnamefont {C.}~\bibnamefont {Friedrich}}, \bibinfo {author}
  {\bibfnamefont {A.~I.}\ \bibnamefont {Lichtenstein}}, \bibinfo {author}
  {\bibfnamefont {M.~I.}\ \bibnamefont {Katsnelson}}, \ and\ \bibinfo {author}
  {\bibfnamefont {S.}~\bibnamefont {Bl\"ugel}},\ }\href {\doibase
  10.1103/PhysRevLett.106.236805} {\bibfield  {journal} {\bibinfo  {journal}
  {Phys. Rev. Lett.}\ }\textbf {\bibinfo {volume} {106}},\ \bibinfo {pages}
  {236805} (\bibinfo {year} {2011})}\BibitemShut {NoStop}%
\bibitem [{\citenamefont {Emelyanenko}\ and\ \citenamefont
  {Boinovich}(2008)}]{kappaorig}%
  \BibitemOpen
  \bibfield  {author} {\bibinfo {author} {\bibfnamefont {A.}~\bibnamefont
  {Emelyanenko}}\ and\ \bibinfo {author} {\bibfnamefont {L.}~\bibnamefont
  {Boinovich}},\ }\href {http://stacks.iop.org/0953-8984/20/i=49/a=494227}
  {\bibfield  {journal} {\bibinfo  {journal} {Journal of Physics: Condensed
  Matter}\ }\textbf {\bibinfo {volume} {20}},\ \bibinfo {pages} {494227}
  (\bibinfo {year} {2008})}\BibitemShut {NoStop}%
\bibitem [{\citenamefont {Dresselhaus}\ and\ \citenamefont
  {Dresselhaus}(1980)}]{dresselhaus}%
  \BibitemOpen
  \bibfield  {author} {\bibinfo {author} {\bibfnamefont {M.~S.}\ \bibnamefont
  {Dresselhaus}}\ and\ \bibinfo {author} {\bibfnamefont {G.}~\bibnamefont
  {Dresselhaus}},\ }\href@noop {} {\bibfield  {journal} {\bibinfo  {journal}
  {Advnaces in Physics}\ }\textbf {\bibinfo {volume} {30}},\ \bibinfo {pages}
  {1} (\bibinfo {year} {1980})}\BibitemShut {NoStop}%
\bibitem [{\citenamefont {Taft}\ and\ \citenamefont {Philipp}(1965)}]{taft}%
  \BibitemOpen
  \bibfield  {author} {\bibinfo {author} {\bibfnamefont {E.~A.}\ \bibnamefont
  {Taft}}\ and\ \bibinfo {author} {\bibfnamefont {H.~R.}\ \bibnamefont
  {Philipp}},\ }\href {\doibase 10.1103/PhysRev.138.A197} {\bibfield  {journal}
  {\bibinfo  {journal} {Phys. Rev.}\ }\textbf {\bibinfo {volume} {138}},\
  \bibinfo {pages} {A197} (\bibinfo {year} {1965})}\BibitemShut {NoStop}%
\bibitem [{\citenamefont {Marinopoulos}\ \emph {et~al.}(2004)\citenamefont
  {Marinopoulos}, \citenamefont {Reining}, \citenamefont {Rubio},\ and\
  \citenamefont {Olevano}}]{mari}%
  \BibitemOpen
  \bibfield  {author} {\bibinfo {author} {\bibfnamefont {A.~G.}\ \bibnamefont
  {Marinopoulos}}, \bibinfo {author} {\bibfnamefont {L.}~\bibnamefont
  {Reining}}, \bibinfo {author} {\bibfnamefont {A.}~\bibnamefont {Rubio}}, \
  and\ \bibinfo {author} {\bibfnamefont {V.}~\bibnamefont {Olevano}},\ }\href
  {\doibase 10.1103/PhysRevB.69.245419} {\bibfield  {journal} {\bibinfo
  {journal} {Phys. Rev. B}\ }\textbf {\bibinfo {volume} {69}},\ \bibinfo
  {pages} {245419} (\bibinfo {year} {2004})}\BibitemShut {NoStop}%
\bibitem [{\citenamefont {Hiraoka}\ \emph {et~al.}(2005)\citenamefont
  {Hiraoka}, \citenamefont {Ishii}, \citenamefont {Jarrige},\ and\
  \citenamefont {Cai}}]{caigraphite}%
  \BibitemOpen
  \bibfield  {author} {\bibinfo {author} {\bibfnamefont {N.}~\bibnamefont
  {Hiraoka}}, \bibinfo {author} {\bibfnamefont {H.}~\bibnamefont {Ishii}},
  \bibinfo {author} {\bibfnamefont {I.}~\bibnamefont {Jarrige}}, \ and\
  \bibinfo {author} {\bibfnamefont {Y.~Q.}\ \bibnamefont {Cai}},\ }\href@noop
  {} {\bibfield  {journal} {\bibinfo  {journal} {Phys. Rev. B}\ }\textbf
  {\bibinfo {volume} {72}},\ \bibinfo {pages} {075103} (\bibinfo {year}
  {2005})}\BibitemShut {NoStop}%
\bibitem [{\citenamefont {Pines}(1999)}]{pines}%
  \BibitemOpen
  \bibfield  {author} {\bibinfo {author} {\bibfnamefont {D.}~\bibnamefont
  {Pines}},\ }\href@noop {} {\emph {\bibinfo {title} {Elementary Excitations in
  Solids}}}\ (\bibinfo  {publisher} {Westview Press},\ \bibinfo {year}
  {1999})\BibitemShut {NoStop}%
\end{thebibliography}%
\end{document}